\documentclass[aps,prd,amssymb,final,nobibnotes,nofootinbib,twocolumn,superscriptaddress]{revtex4-2}
\usepackage{graphicx}
\usepackage{hyperref}
\usepackage{amsmath}
\usepackage{color}
\usepackage{xcolor}
\usepackage{numprint}
\usepackage{float}
\usepackage{ulem}
\usepackage{physics}
\usepackage{mathtools}
\usepackage{dsfont}
\usepackage{subfigure}
\usepackage{upgreek}
\usepackage{multirow}
\usepackage{mathrsfs}
\usepackage{comment}

\definecolor{darkgreen}{rgb}{0,0.5,0}

\newcommand{\ii}{\mathrm{i}}
\newcommand{\eq}[1]{\eqref{#1}}

\begin{document}
\title{Interaction of the gravitational Hawking radiation and a static point mass}

% \author{Author 1}
% \email{e-mail 1} 
% \affiliation{affiliation a}

% \author{Author 2}
% \email{e-mail 2} 
% \affiliation{affiliation b}

% \author{Author 3}
% \email{e-mail 3} \affiliation{affiliation c}

\author{Jo\~ao P. B. Brito}
\email{joao.brito@icen.ufpa.br} 
\affiliation{Programa de P\'os-Gradua\c{c}\~{a}o em F\'{\i}sica, Universidade Federal do Par\'a, 66075-110, Bel\'em, Par\'a, Brazil}
\affiliation{Department of Mathematics,  University of York,  Ian Wand Building, Deramore Lane, York YO10 5GH, United Kingdom}
%\affiliation{Department of Mathematics,  University of York,  Heslington, York YO105DD, United Kingdom}
		
\author{Atsushi Higuchi}
\email{atsushi.higuchi@york.ac.uk}
\affiliation{Department of Mathematics,  University of York,  Ian Wand Building, Deramore Lane, York YO10 5GH, United Kingdom}
%\affiliation{Department of Mathematics,  University of York,  Heslington, York YO105DD, United Kingdom}

\author{Lu\'is C. B. Crispino}
\email{crispino@ufpa.br}
\affiliation{Programa de P\'os-Gradua\c{c}\~{a}o em F\'{\i}sica, Universidade Federal do Par\'a, 66075-110, Bel\'em, Par\'a, Brazil}

\date{\today}
\begin{abstract}
 We study the interaction of a stress-energy tensor describing a static point mass supported by a string outside a Schwarzschild black hole with the gravitons of the Hawking radiation. We derive a closed-form analytic expression for the total response rate of  
 this stress-energy tensor to the thermal gravitons in the Unruh state, which models the quantum state in the spacetime of 
 a spherically symmetric black hole formed by gravitational collapse. 
This response rate is finite  
in contrast with the infrared divergent response rate for a static point mass supported by a string in Rindler spacetime, i.e., a point mass
accelerated uniformly by a string in Minkowski spacetime. 
By comparing the response rate near the black hole horizon with that in Rindler spacetime, we show that the size of the black hole  
acts as a natural infrared cutoff. We also find that the response rate 
of this stress-energy tensor to the thermal gravitons incoming from past null infinity
in the Hartle-Hawking state vanishes.  As a result, the total response rate of a static point mass (supported by a string)
in the Unruh and Hartle-Hawking states for gravitons are identical.  This is also the case for a static charge interacting with
the electromagnetic field but not for a static source for a massless scalar field.

\smallskip

%\noindent
%\textbf{Keywords:} Gravitational perturbations, radiation, bremsstrahlung, response rate
\end{abstract}

\maketitle

%%%%%%%%%%%%%%%%%%%%%%%%%%%%%
\section{Introduction}
%%%%%%%%%%%%%%%%%%%%%%%%%%%%%
It is widely believed that a theory of quantum gravity is required to describe the earliest moments of the Universe and the final stages of black hole evaporation when quantum effects of gravity are expected to dominate. Despite substantial efforts toward its formulation, a fully consistent and well-defined theory of quantum gravity remains as one of the major open problems in fundamental physics (see Ref.~\cite{DeWitt_2008} for a list of proposed approaches). In the absence of such a fully consistent theory, one is led to semiclassical frameworks that are believed to delineate the interface between classical and quantum descriptions of the underlying gravitational interaction. In particular, quantum field theory in curved spacetime provides a well-established setting that consistently combines quantum field theory with general relativity at the perturbation level, describing quantum fields propagating on a fixed classical background. One of the most remarkable predictions of this framework is the Hawking effect, according to which black holes emit a thermal flux of particles, leading to their gradual evaporation~\cite{hawking_1974,hawking_1975}. By predicting that black holes emit this thermal flux of particles with a characteristic temperature, the Hawking effect played a central role in connecting black holes to thermodynamics, leading to their interpretation as thermodynamic systems. 

A phenomenon  closely related to the Hawking effect is the Unruh effect,  
i.e., the fact that the Minkowski vacuum in quantum field theory is perceived as a thermal bath, the Fulling-Davies-Unruh~\cite{fulling_1973,davies_1975,unruh_1976} thermal bath, by uniformly accelerated observers, i.e., by observers in the Rindler frame~\cite{crispino_2008}. The Unruh effect brings into sharp focus the observer-dependent nature of the particle concept in quantum field theory. It is an essential ingredient for a consistent description of quantum field theory from the viewpoint of uniformly accelerated observers~\cite{unruh_wald_1984}. More specifically, it provides the correct vacuum structure and particle interpretation, which are crucial, for example, in the analysis of classical radiation processes described from the perspective of the uniformly accelerated observers~\cite{higuchi_2025}.  
Like the Hawking effect, the Unruh effect has been derived rigorously through several independent approaches, firmly establishing its theoretical validity~\cite{bisognano_1975,bisognano_1976,sewell_1982,fulling_unruh_2004}.

Recently, the present authors 
investigated the gravitational waves emitted by a point mass in uniform acceleration being pulled by a string  
in Minkowski spacetime, which is static 
in the Rindler frame~\cite{brito_2024_gw}. (See also Ref.~\cite{portales-oliva_2024}.) In particular, we computed the response rate of this static, conserved stress-energy tensor 
to the gravitons of the Fulling-Davies-Unruh thermal bath in the co-accelerating frame and showed that it agrees with the graviton emission rate computed in the
standard method in the inertial frame in Minkowski spacetime. We found that the total response rate exhibits an infrared power-law divergence. 

The main purpose of this paper is to investigate the analogous problem of finding  
the response rate of 
a point mass held at rest outside a Schwarzschild black hole by a string to 
the gravitational Hawking radiation, which is modeled by the radiation from the past horizon in the Unruh state~\cite{unruh_1976}. 
For this purpose  
we derive analytic expressions for the graviton mode functions in the low-frequency regime in Schwarzschild spacetime. In particular, we solve the radial equation of motion for the vector-type (odd-parity) modes and apply the Chandrasekhar transformation to obtain the scalar-type (even-parity) modes, which are the ones that couple to the stress-energy tensor considered here.  
We also consider the Hartle-Hawking state as the initial state of the quantum field, which describes a black hole in thermal equilibrium with a surrounding thermal bath at the Hawking temperature~\cite{hartle_1976}.

The remainder of the paper is organized as follows. In Sec.~\ref{sec:stress-energy}, we 
construct a conserved stress-energy tensor satisfying the weak energy condition and corresponding to a point mass at a radial position $r_0$ supported by a one-dimensional radial string extended to infinity.
In Sec.~\ref{sec:perturbations}, we review the gravitational perturbations of the 
Schwarzschild spacetime (Sec.~\ref{sec:even_odd_perturbations}), derive the mode 
solutions analytically in the low-frequency regime 
(Sec.~\ref{sec:gw_ZFL_modes}), and quantize the perturbations using the canonical 
quantization prescription (Sec.~\ref{sec:quantization}).
In Sec.~\ref{sec:response-rate}, we compute the response rate of the conserved 
stress-energy tensor  
constructed in Sec.~\ref{sec:stress-energy} to the thermal bath of the Hawking 
radiation of the Schwarzschild black hole. Our concluding remarks are presented 
in Sec.~\ref{sec:remarks}.
In Appendices~\ref{sec:appendix_A} and~\ref{sec:appendix_B} we discuss technical details that have been omitted from the main text. In Appendix~\ref{sec:appendix_C}, we analyze the response rate of an alternative stress-energy tensor with an additional parameter characterizing the string.
In Appendix~\ref{sec:appendix_D}, we present the response rates of static point particle systems coupled to bosonic fields in some static spacetimes, summarizing the
results of this paper and those in the literature.
We adopt metric signature $(-, +, +, +)$  and units where $G=c=\hbar=1$, unless stated otherwise.

%%%%%%%%%%%%%%%%%%%%%%%%%%%%%
\section{Conserved Stress-Energy Tensor}
\label{sec:stress-energy}
%%%%%%%%%%%%%%%%%%%%%%%%%%%%%
The line element of the Schwarzschild spacetime is given by
\begin{equation}
\label{eq:line_elem}
    \dd\tau^2 = -f(r)\dd t^2 + \frac{\dd r^2}{f(r)} + r^2(\dd\theta^2+\sin^2\theta\,\dd\phi^2),
\end{equation}
where the Schwarzschild radial 
function is
\begin{equation}
\label{eq:Swarzschild}
f(r) = 1-\frac{r_h}{r},
\end{equation}
with $r_h \equiv 2M$ being the radial coordinate of the event horizon, where $M$ is the mass of the black hole.

We begin by constructing a conserved stress-energy tensor, $T^{\mu\nu}$, in the spacetime described by Eqs.~\eqref{eq:line_elem} and \eqref{eq:Swarzschild}, for a 
system  
that includes a static point mass located at a fixed radial position, $r_0 > r_h$. A particle at rest at fixed $r_0$ outside the black hole does not follow a geodesic and therefore  
has a nonvanishing proper acceleration,
$a^\mu = u^\nu \nabla_\nu u^\mu \neq 0$, where $u^\mu$ denotes the four-velocity. 
This implies that the particle undergoes uniform acceleration in the exterior region of the black hole. The agent responsible for supporting this constant 
acceleration, which continuously exchanges energy-momentum with the particle, must therefore be  
included. 
Thus, only the 
stress-energy tensor of the whole system, i.e., the point mass together 
with the agent driving the acceleration, is conserved. (This conservation is essential for ensuring the gauge 
invariance.)

We assume that the only nonvanishing components of the stress-energy tensor are 
$T^{tt}$ and $T^{rr}$, and that they are proportional to
$\delta_{\perp}^{(2)} \coloneqq 
\delta(\theta-\theta_0)\delta(\phi-\phi_0)/\sin\theta$.
 The $r$-component of the conservation equation, $\nabla_\mu T^{\mu \nu}=0$, gives
\begin{equation}
    \label{eq:r-comp_conserv_eq}
     \partial_r T^{rr} + \left(\frac{2}{r}-\frac{f'}{2f}\right)T^{rr} + \frac{f f'}{2}\,T^{tt}=0.
\end{equation} 
We let the $T^{rr}$ component be given by
\begin{equation}
    \label{eq:ansatz_for_Trr}
    T^{rr} = -\mu\,\frac{f}{r^2}\,F(r)\,\delta_\perp^{(2)},
\end{equation}
with $F$ being a real-valued and positive function of $r$ and $\mu = \mu (r_0)> 0$ is related to the point particle mass, $m_0$, as shown bellow. Substituting  
Eq~\eqref{eq:ansatz_for_Trr} into Eq. \eqref{eq:r-comp_conserv_eq}, we obtain
\begin{equation}
    \label{eq:tt_comp_of_Tmunu}
    T^{tt}= \frac{\mu}{r^2} \left[\frac{2}{f'(r)} F'(r)+\frac{1}{f(r)}F(r)\right]\, \delta_{\perp}^{(2)}.
\end{equation}
The stress-energy tensor given by Eqs. \eqref{eq:ansatz_for_Trr} and \eqref{eq:tt_comp_of_Tmunu} satisfies the weak energy condition provided that $F'(r) \geq 0$. 

\begin{figure}
\center
\includegraphics[width=.92 \linewidth]{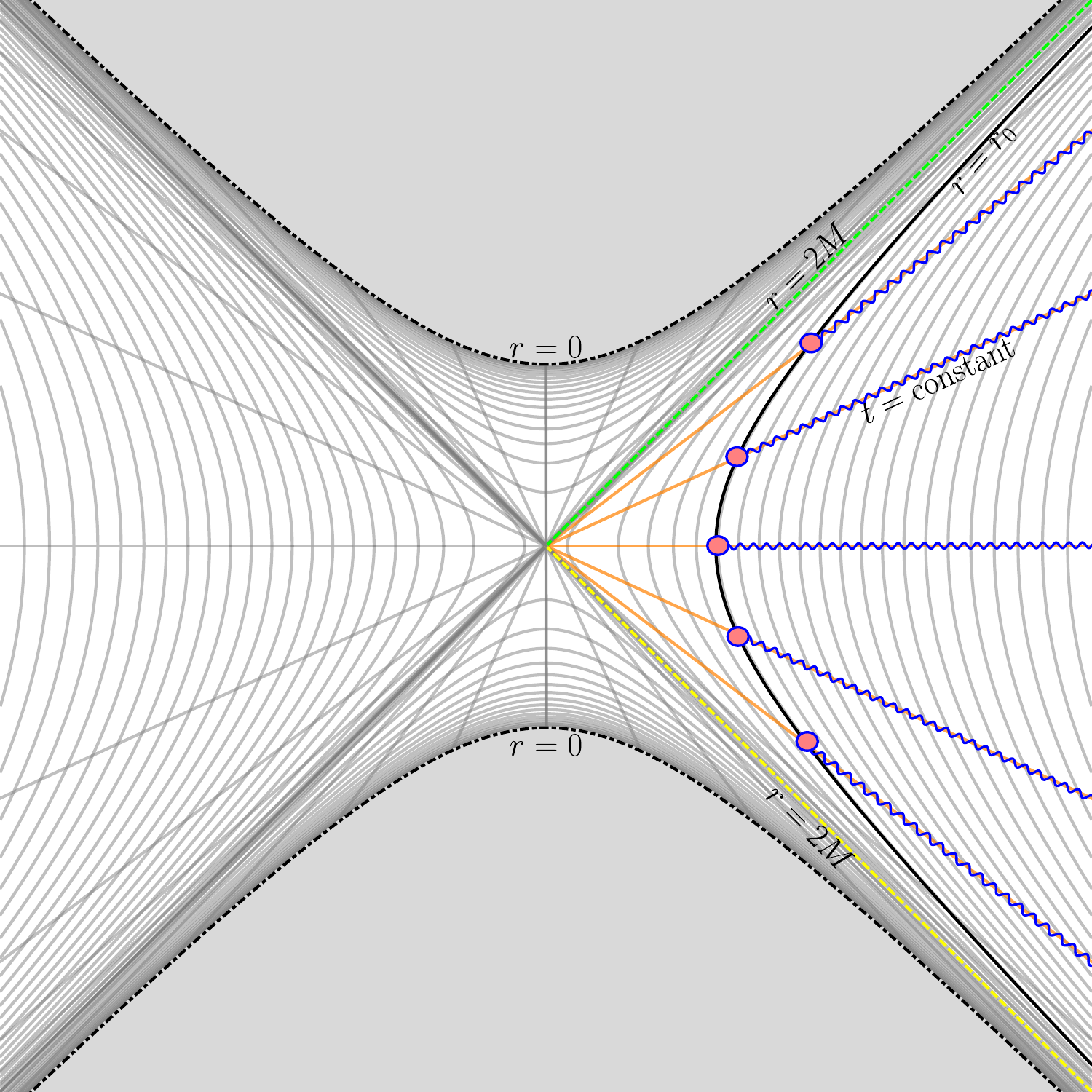}
\caption{Kruskal-Szekeres diagram of Schwarzschild spacetime illustrating the 
stress-energy tensor given by Eqs.~\eqref{eq:T_t_t} and 
\eqref{eq:T_r_r}. The point mass located at $r = r_0$, supported by a string 
extending to $r > r_0$, is shown in the exterior region of the black hole as dots 
with wavy tails on each $t=\mathrm{constant}$ slices.
}
\label{fig:Schwarzschild_spacetime}
\end{figure}

Now, we  
choose the function $F(r)$ as
\begin{equation}
F(r) = \theta (r-r_0),
\label{eq:F}
\end{equation}
where $\theta(x)$ is the Heaviside step function. This  
choice leads to the following stress-energy tensor: 
\begin{equation}
\label{eq:T_t_t}
T^{tt} = \frac{\mu}{r^2} \left[\frac{2}{f'(r)} \delta(r-r_0)+\frac{1}{f(r)}\theta(r-r_0) \right]  
\delta_{\perp}^{(2)},
\end{equation}
\begin{equation}
\label{eq:T_r_r}
    T^{rr} = -\mu\,\theta(r-r_0)\,\frac{f(r)}{r^2}\,\delta_\perp^{(2)},
\end{equation}
with all the other tensor components vanishing. This stress-energy tensor describes a classical system consisting of a point mass static outside the black hole at $r=r_0$ being supported by a string extended to $r>r_0$. Figure~\ref{fig:Schwarzschild_spacetime} illustrates this system in the Kruskal-Szekeres diagram of 
the
Schwarzschild spacetime. The classical system is represented as dots with wavy tails on five $t= \text{constant}$ slices of the diagram.

To find the relation between $\mu(r_0)$ and $m_0$, we first note that the stress-energy tensor of a point particle with mass $m_0$ and world line $z^\mu(\tau)$, where $\tau$ is its proper time, is given by
\begin{equation}
    \label{eq:point_mass_Tmunu}
     T^{\mu\nu}_{p}
    = m_0 \int \frac{u^\mu u^\nu}{\sqrt{-g}}\,\delta^{(4)}\!\bigl(x-z(\tau)\bigr)\,\dd\tau.
\end{equation}
For a static point mass at $r=r_0$, the four-velocity is given by $u^\mu=(u^t,0,0,0)$, with $u^t=1/\sqrt{f(r_0)}$. Substituting this into Eq.~\eqref{eq:point_mass_Tmunu}, we obtain, noting that $\delta(t-t(\tau)) = \sqrt{f(r_0)}\delta(\tau-\tau(t))$,
\begin{equation}
    T^{tt}_{p}(x)=m_0\frac{1}{r_0^2\sqrt{f(r_0)}}\,\delta(r-r_0)\,\delta_\perp^{(2)}.
    \label{eq:point-mass}
\end{equation}
By comparing Eq.~\eqref{eq:point-mass} with the first term proportional to the $\delta$-function in 
Eq.~\eqref{eq:T_t_t}, we find that
\begin{eqnarray}
    \label{eq:mu}
    \mu(r_0) &=& \frac{1}{2} m_0 f'(r_0) f(r_0)^{-1/2} \\
        \label{eq:mu_a}
    &=& m_0 \varrho(r_0),
\end{eqnarray}
which is the ``weight'' of the point mass as measured by a static observer at $r_0$, where 
\begin{equation}
    \label{eq:acceleration} 
    \varrho(r_0) = \frac{1}{2} f'(r_0) f(r_0)^{-1/2}
\end{equation}
is the magnitude of the proper four-acceleration, $\varrho^\mu$, of the point mass, which is given by $\varrho^\mu \equiv u^\nu \nabla_\nu u^\mu$ and $\varrho \equiv \sqrt{\varrho^\mu \varrho_\mu}$.
Note that the stress-energy tensor~\eqref{eq:point_mass_Tmunu} satisfies the conservation equation, $\nabla_\mu T^{\mu\nu}_{p} = 0$, if an only if the $4$-vector $u^\mu$ obeys the geodesic equation, $u^\nu \nabla_\nu u^\mu = 0$.  This is why we need the contribution from the string supporting the point mass.
From Eqs.~\eqref{eq:T_t_t} and \eqref{eq:T_r_r}, we see that the stress-energy tensor diverges in the limit $r_0 \to r_h$,
as the force $m_0\varrho(r_0)$ required to hold it static at that radial position diverges.  

As expected, near the black hole horizon, the stress-energy tensor \eqref{eq:T_t_t}--\eqref{eq:T_r_r} reduces to the corresponding expression in Rindler spacetime~\cite{brito_2024_gw}. To see this, note that
$f(r) \approx f'(r_h)(r-r_h)$ near the horizon,
with $f'(r_h) \approx 1/r_h$. 
We assume that $r_0 - r_h \ll r_h$ and  
change the radial coordinate from $r$ to $\xi$ with the relation
\begin{equation}
\mathrm{e}^{2 a\xi} = \frac{r-r_h}{r_0-r_h},
\end{equation}
where $a = \varrho(r_0)\approx \frac{1}{2}f'(r_h)^{1/2}(r_0-r_h)^{-1/2}$
and hence $\mu\approx m_0 a$. Note that $\xi=0$ at $r=r_0$.  By defining
$\eta := \sqrt{f(r_0)}\,t$, which is the proper time at $r=r_0$ ($\xi=0$), the line element near the horizon is approximated as
\begin{equation}
    \dd \tau^2 \approx \mathrm{e}^{2a\xi}(-\mathrm{d}\eta^2 +\mathrm{d}\xi^2) + r_h^2(\mathrm{d}\theta^2 + \sin^2\theta
    \mathrm{d}\phi^2)\,,
\end{equation}
which is approximately the standard metric of Rindler spacetime.  
Rewriting the
stress-energy tensor in terms of $m_0$, $a$, and the Rindler coordinates $(\eta,\xi)$ defined above, we obtain
\begin{eqnarray}
    \label{eq:T_tt_Rindler}
    T^{\eta\eta} & \approx &   
m_0\left[ \delta(\xi) + 
a\, \mathrm{e}^{-2 a \xi}\, \theta(\xi)\right]\frac{1}{r_{h}^{2}}\,\delta_{\perp}^{(2)}, \\
    \label{eq:T_rr_Rindler}
    T^{\xi\xi} & \approx &
-m_0 a\, \mathrm{e}^{-2a\xi}\,\theta(\xi)
\frac{1}{r_{h}^{2}}\,\delta_{\perp}^{(2)}, 
\end{eqnarray}
which coincide with the components of the stress-energy tensor obtained in Ref.~\cite[Eqs.~(8)--(9)]{brito_2024_gw}. 

%%%%%%%%%%%%%%%%%%%%%%%%%%%%%
\section{Gravitational Perturbations in Schwarzschild spacetime and their Quantization}
\label{sec:perturbations}
%%%%%%%%%%%%%%%%%%%%%%%%%%%%%
In this section, we  
present the relevant gravitational perturbations in Schwarzschild spacetime,  
which can be found, e.g., in Refs.~\cite{kodama_2000,kodama_2003,bernar_2017}. We also derive analytic expressions for the gravitational modes in the low-frequency regime by explicitly solving the Regge-Wheeler equation for the vector-type modes and using the Chandrasekhar transformation to obtain the Zerilli modes, i.e., the scalar-type modes, in this regime. Then, we quantize the gravitational perturbations following the canonical quantization prescription. 

%%%%%%%%%%%%%%%%%%%%%%%%%%%%%%%%%%%%
\subsection{Scalar-type and vector-type gravitational perturbations}
\label{sec:even_odd_perturbations}
%%%%%%%%%%%%%%%%%%%%%%%%%%%%%%%%%%%%

The scalar-type gravitational perturbations in Schwarzschild spacetime are 
given as
\begin{align}
h^{(S,n; \omega \ell m)}_{ai} &= 0, \label{eq:scalar_ai}\\
h^{(S,n; \omega \ell m)}_{ij} &= 2r^2 \gamma_{ij} F^{(n;\ell)} Y_{\ell m}, \label{eq:scalar_ij}\\
h^{(S,n; \omega \ell m)}_{ab} &= F^{(n;\ell)}_{ab} Y_{\ell m}, \label{eq:scalar_ab}
\end{align}
where the indices $a,b$ denote components on the $(t,r)$ submanifold (orbit space), and $i,j$ label those on the 
unit two-sphere ($S^2$); 
$\gamma_{ij}$ is the metric of $S^{2}$. The scalar spherical harmonics $Y_{\ell m}$ are 
given in terms of the associated Legendre  
functions, $P_\ell^m(\cos \theta)$, as~\cite{nist} 
\begin{equation}
    \label{eq_spherical}
 Y_{\ell m}(\theta, \phi) = \sqrt{\frac{2\ell + 1}{4\pi} \frac{(\ell - m)!}{(\ell + m)!}} \, P_{\ell}^{m}(\cos\theta) \, \mathrm{e}^{\ii m \phi},
\end{equation}
and the functions $F^{(n;\ell)}$ and $F^{(n;\ell)}_{ab}$ are expressed in terms of a master variable $\Phi^{S}_{\ell}(t,r)$ as
\begin{align}
F^{(n;\ell)} &= \frac{1}{4r^2 H_\ell}\Big[ (H_\ell - r f')\,\Omega^S_{\omega \ell} + 2r D^a r D_a \Omega^S_{\omega \ell} \Big], \label{eq:F_scalar}\\
F^{(n;\ell)}_{ab} &= \frac{1}{H_\ell}\!\left(D_a D_b - \tfrac{1}{2} g_{ab} \Box \right) \!\Omega^S_{\omega \ell}, \label{eq:Fab_scalar}
\end{align}
with
\begin{eqnarray}
\label{eq:H_def}
H_\ell &=& \frac{1}{r} \left[(\Lambda_\ell-2)r + 6M \right], \\
\label{eq:Omega_def}
\Omega^S_{\omega \ell} &=& r H_\ell \Phi^S_{\omega \ell}, \\
\label{eq:Lambda_l}
\Lambda_\ell &=& \ell(\ell+1).
\end{eqnarray}
In Eqs.~\eqref{eq:F_scalar} and \eqref{eq:Fab_scalar}, $D_a$ is the covariant derivative on the two-dimensional orbit space, and $\Box$ is the d'Alambertian in the orbit space, given by
\begin{equation}
    \label{eq:dalambertian}
    \Box \equiv -f(r)^{-1} \frac{\partial^{2}}{\partial t^{2}}
+ \frac{\partial}{\partial r} \!\left[ f(r)\, \frac{\partial}{\partial r} \right].
\end{equation}
%\tb{Note that the operator $D_a D_b - \tfrac{1}{2} g_{ab} \Box$ in Eq.~\eqref{eq:Fab_scalar} is trace-free.}

The vector-type gravitational perturbations in Schwarzschild spacetime are written is terms of a master variable $\Phi^{V}_{\ell}$ as
\begin{equation}
    \label{eq:vector_perturba}
    h^{(V,n; \omega \ell m)}_{a i}
= Y^{(\ell m)}_{i}\, \epsilon_{a b}\, D^{b}\!\left( r\, \Phi^{V}_{\ell} \right),
\end{equation}
with all other components vanishing. Here, $\epsilon_{a b}$ is the totally antisymmetric tensor (Levi-Civita tensor) on the orbit space, with the components defined by $\epsilon_{tt} = \epsilon_{rr}=0$ and $\epsilon_{tr} = - \epsilon_{rt} = 1$. In Eq.~\eqref{eq:vector_perturba}, $Y^{(\ell m)}_{i} = Y^{(\ell m)}_{i}(\theta, \phi)$, with $i \in \left\{ \theta, \phi \right\}$, are the vector spherical harmonics~\cite{nist,higuchi_1987,higuchi_1987b}. These functions are expressed as
\begin{equation}
    \label{eq_vec_spherical}
Y_{i}^{(\ell m)}(\theta, \phi) =  \frac{1}{\sqrt{\Lambda_\ell}} \epsilon_{i j} \, \partial^{j} Y_{\ell m}(\theta, \phi), \quad (\ell \geq 1),
\end{equation}
where $\epsilon_{ij}$ is the totally antisymmetric tensor (Levi-Civita tensor) on $S^2$ 
(with indices representing the coordinates $\theta$ and $\phi$), with the components defined by $\epsilon_{\theta\theta} = \epsilon_{\phi\phi} = 0$ and $\epsilon_{\theta\phi} = -\epsilon_{\phi\theta} = \sin\theta$.

The positive-frequency master functions are written as
\begin{equation}
\label{eq:master_and_u}
\Phi^{P}_{\omega \ell}(t,r) = u_{\omega\ell}^{P}(r)\,\mathrm{e}^{-\ii \omega t}, \quad \omega > 0,
\end{equation}
with $P \in \{S,V\}$ and the functions $u_{\omega\ell}^{P}(r)$ satisfying the following homogeneous differential equation,
\begin{equation}
    \label{eq:schrodinger_eq}
    \left[\dv[2]{}{x}
+  \omega^{2} - V_{P}(r) \right] u^{P}_{\omega \ell}(r) = 0,
\end{equation}
where the tortoise coordinate, $x$, is given by 
\begin{equation}\label{eq:tortoise}
    x(r) = r+2M\ln\left(\frac{r}{2M} -1\right).
\end{equation} 
In Eq.~\eqref{eq:schrodinger_eq}, $V_{P}(r)$ denotes the Zerilli potential, $V_S(r)$, and the Regge-Wheeler potential, $V_V(r)$. They are 
given by~\cite{chandrasekhar_1983}
\begin{eqnarray}
\label{eq:zerilli_pot}
V_S(r) &=& +\beta \frac{\dd g}{\dd x} + \beta^2 g^2 + k_\ell g,\\
\label{eq:regge-wheeler_pot}
V_V(r) &=& -\beta \frac{\dd g}{\dd x} + \beta^2 g^2 + k_\ell g,
\end{eqnarray}
where $\beta \equiv 6M$,
\begin{equation}
    \label{eq:k}
    k_\ell \equiv \Lambda_\ell (\Lambda_\ell -2),
\end{equation}
and the function $g(r)$ is given by
\begin{equation}
    \label{eq:g}
    g(r) \equiv \frac{f(r)}{r^2 H_\ell}.
\end{equation}
The potential $V_V(r)$ can be simplified as
\begin{equation}
\label{eq:regge_wheeler_pot}
V_V(r) = f(r)\left(\frac{\Lambda_\ell}{r^2} - \frac{6M}{r^3}\right).    
\end{equation}

\begin{widetext}
The potentials $V_P(r)$, given by Eqs.~\eqref{eq:zerilli_pot} and~\eqref{eq:regge-wheeler_pot}, vanish in the limit $r\to r_h$ ($x\to -\infty$) faster than any powers of $1/|x|$ as $V_P \sim \exp\left(-\abs{x}/r_h\right)$, and 
is approximately $\Lambda_\ell/r^2$  
for large $r$. Thus, we  
have the independent solutions, $u^{P,\text{in}}_{\omega \ell}$ and  $u^{P,\text{up}}_{\omega \ell}$, to  Eq.~\eqref{eq:schrodinger_eq}
satisfying the following boundary conditions:
\begin{equation}
\left(A^{P}_{\omega \ell}\right)^{-1} u^{P,\text{in}}_{\omega \ell} \approx
\begin{cases}
 \mathcal{T}^{P,\text{in}}_{\omega \ell} \mathrm{e}^{- \ii \omega x}, & (x<0, \abs{x} \gg r_h), \\
 (-\ii)^{\ell+1} \omega \, x \, \overline{h_\ell^{(1)}(\omega x)} +  \ii^{\ell+1} \mathcal{R}^{P,\text{in}}_{\omega \ell}\omega\, x \, h_\ell^{(1)}(\omega x), & (x \gg r_h),
\end{cases}
\label{eq:Rin_asymp}
\end{equation}
and
\begin{equation}
\left(A^{P}_{\omega \ell}\right)^{-1} u^{P,\text{up}}_{\omega \ell} \approx
\begin{cases} \mathrm{e}^{\ii \omega x} + \mathcal{R}^{P,\text{up}}_{\omega \ell} \mathrm{e}^{- \ii \omega x}, & (x<0, \abs{x} \gg r_h), \\
 \ii^{\ell +1}\mathcal{T}^{P,\text{up}}_{\omega \ell} \omega \, x \, h_\ell^{(1)}(\omega x), & (x \gg r_h),
\end{cases}
\label{eq:Rup_asymp}
\end{equation}
where the overline denotes complex conjugation, $A^{P}_{\omega \ell}$ denotes the overall normalization constants to be determined, and $h_\ell^{(1)}$ denotes the spherical Bessel functions of the third kind~\cite{nist}. The modes labeled ``in'' are the ones purely incoming from past null infinity $\mathscr{I}^-$; the modes labeled ``up'' are the ones purely incoming from the past black hole horizon $\mathscr{H}^{-}$. The transmission coefficients are
$|\mathcal{T}^{P,n}_{\omega\ell}|^2$ whereas the reflection coefficients are $|\mathcal{R}^{P,n}_{\omega\ell}|^2$.
\end{widetext}

The static stress-energy tensor interacts with the zero-energy gravitons of the Hawking radiation. We next derive analytic expressions for the gravitational perturbation modes in the low-frequency regime.

%%%%%%%%%%%%%%%%%%%%%%%%%%%%%
\subsection{Low-frequency regime of the vector-type and scalar-type gravitational modes}
\label{sec:gw_ZFL_modes}
%%%%%%%%%%%%%%%%%%%%%%%%%%%%%
We first rewrite the differential equation \eqref{eq:schrodinger_eq} with $\omega=0$ for the vector-type modes, i.e., the modes associated with the potential given by Eqs.~\eqref{eq:regge-wheeler_pot} and~\eqref{eq:regge_wheeler_pot}, by setting $u^{V}_{\omega \ell}(r) = r Z_\ell(r)$. Defining a new variable as $z \coloneqq r/M-1$, this  
differential equation becomes
\begin{equation}
\label{eq:Z-equation}
(1-z^2)Z_\ell'' - 2zZ_\ell' + \left(\Lambda_\ell - \frac{\rho_s}{z+1}\right) Z_\ell = 0,
\end{equation}
where $\rho_s \equiv 8$.\footnote{The differential equation governing a field of spin $s$ takes the form of Eq.~\eqref{eq:Z-equation} with $\rho_s \coloneqq 2s^{2}$, where $s=0$ for a scalar field, $s=1$ for the electromagnetic field, and $s=2$ for the vector-type gravitational perturbations.}
Equation \eqref{eq:Z-equation}  
would be the Legendre differential equation 
if the term proportional to $(z+1)^{-1}$ were absent. 
The general solution to this equation, written in terms of the Legendre functions of the first and second kind,  
$P_{\ell}(z)$ and $Q_\ell (z)$, respectively, is given by
\begin{equation}
        \label{eq:ZFL_regge-wheeler_sol}
    Z_\ell(z) = 
    c_1 \widehat{J}_\ell P_\ell(z) +c_2 \widehat{J}_\ell Q_\ell(z),\ \ c_1,c_2\in \mathbb{C},
\end{equation}
where $\widehat{J}_\ell$ denotes the differential operator
\begin{equation}
    \label{eq:J_ell}
    \widehat{J}_\ell = 1 - 4\frac{(z-1)}{\Lambda_\ell}\dv{\,}{z}
         + 6\frac{(z-1)^2}{k_\ell}\dv[2]{\,}{z},
\end{equation}
with $k_\ell$ and $\Lambda_\ell$ given by Eqs.~\eqref{eq:k} and \eqref{eq:Lambda_l}, respectively. 
First, we note that $P_\ell(z) \sim z^\ell$ and $Q_\ell(z) \sim z^{-\ell-1}$ for $z \gg 1$, while $P_\ell(z) \approx 1$ and $Q_\ell(z) \approx -\ln\sqrt{z-1}$ for $z \approx 1$.  
In the $\omega \to 0$ limit, the transmission coefficients in Eqs.~\eqref{eq:Rin_asymp} and \eqref{eq:Rup_asymp} tends to $0$. Thus, the in-modes, which describe gravitational waves incoming from $\mathscr{I}^-$, should not diverge
as $z\to 1$ ($r\to 2M$), whereas the up-modes, which describe gravitational waves coming from the past horizon, $\mathscr{H}^{-}$, should vanish in the limit $z\to \infty$ ($r\to\infty$).
Thus, in the $\omega\to 0$  limit the in-modes should be identified with the solution given in terms of $P_\ell(z)$ 
whereas the up-modes should be identified with those given in terms of 
$Q_\ell(z)$. 
The in-mode and up-mode radial solutions, $u^{V,\text{in}}_{\omega \ell}$ and $u^{V,\text{up}}_{\omega \ell}$, for $\omega \approx 0$ are then approximated as
\begin{equation}
    \label{eq:u_in_ZFL_sol}
     u_{\omega \ell}^{V,\text{in}} \approx C_{\omega \ell}^{\text{in}}\, r \widehat{J}_\ell P_\ell\left(\frac{r}{M} -1\right),
\end{equation}
\begin{equation}
    \label{eq:u_up_ZFL_sol}
    u_{\omega \ell}^{V,\text{up}} \approx C_{\omega \ell}^{\text{up}}\, r \widehat{J}_\ell Q_\ell\left(\frac{r}{M} -1\right),
\end{equation}
where $C^{\text{in}}_{\omega \ell}$ and 
$C^{\text{up}}_{\omega \ell}$ are normalization constants to be determined.

We first determine the constant $C^{\text{up}}_{\omega \ell}$. Substituting the approximate formula $Q_\ell(z) \approx - (1/2) \ln (z-1)$ near the horizon ($z\approx 1$)  
into Eq.~\eqref{eq:u_up_ZFL_sol}, we find for negative and large $x$
\begin{equation}
    \label{eq:u_up_ZFL_sol_near_horizon}
    u_{\omega \ell}^{V,\text{up}} \approx -\frac{C_{\omega \ell}^{\text{up}}}{2}\,x + K,
\end{equation}
for some constant $K$,
where we have used the near-horizon approximation $x \approx r_h \ln(z - 1)$ for the tortoise coordinate~\eqref{eq:tortoise}. 
On the other hand, 
by using the approximation $\mathrm{e}^{\pm \ii \omega x} \approx 1 \pm \ii \omega x$ for $r_h \ll -x \ll 1/\omega$ in Eq.~\eqref{eq:Rup_asymp}, we obtain
\begin{equation}
    \label{eq:Rup_asymp_near_horizon}
     u_{\omega \ell}^{P,\text{up}} \approx A^{P}_{\omega \ell}\left[\left(1+\mathcal{R}^{P,\text{up}}_{\omega \ell} \right) + \ii \omega x \left(1-\mathcal{R}^{P,\text{up}}_{\omega \ell} \right)\right].
\end{equation}
Since the mode has total reflection in the limit $\omega\to 0$, we have $|\mathcal{R}^{P,\text{up}}_{\omega \ell}|\to 1$ in this limit. Then, for Eqs.~\eqref{eq:u_up_ZFL_sol_near_horizon} and \eqref{eq:Rup_asymp_near_horizon}
to be consistent with each other, we must have $1+\mathcal{R}^{P,\text{up}}_{\omega \ell} \to 0$ as
$\omega\to 0$ and
\begin{equation}
    \label{eq:Din}
C_{\omega \ell}^{\text{up}} = - 4\, \ii \, \omega A^{V}_{\omega \ell},
\end{equation}
for $\omega\approx 0$. 
The constant $C_{\omega \ell}^{\text{in}}$ can be determined   
by a similar argument. Note first
\begin{equation}
    \label{eq:legendreP_asymp}
    P_\ell(z) \approx \frac{(2\ell)!}{2^\ell (\ell!)^2} z^\ell, \quad \text{for} \,\, z \gg 1,
\end{equation}
which follows immediately from the well-known formula,
\begin{equation}
    P_\ell(z) = \frac{1}{2^\ell \ell!}\frac{\dd^{\ell}\,}{\dd z^\ell}(z^2-1)^{\ell}.
\end{equation}
Then, by substituting Eq.~\eqref{eq:legendreP_asymp} into Eq.~\eqref{eq:u_in_ZFL_sol} we find
\begin{equation}
\label{eq:u_in_ZFL_asymp_low_omega}
    u_{\omega \ell}^{V,\text{in}} \approx \, C_{\omega \ell}^{\text{in}} \,\frac{(2\ell)!}{2^\ell (\ell+2)!(\ell-2)!}\frac{r^{\ell+1}}{M^\ell}, \quad \mathrm{for\,\,} r \gg r_h.
\end{equation}
 Now, the approximate expression of $u_{\omega \ell}^{V,\text{in}}$  
for $r \gg r_h$ in Eq.~\eqref{eq:Rin_asymp} can be written using $h^{(1)}_\ell(\rho) = j_\ell(\rho) +\ii\, y_\ell(\rho)$ as
\begin{align}
\label{eq:u-in-j-and-n}
    u_{\omega \ell}^{P,\text{in}} & \approx A_{\omega\ell}^P\large\{ (-\ii)^{\ell+1}\omega x
    [ 1 + (-1)^{\ell+1}\mathcal{R}_{\omega\ell}^P]j_\ell(\omega x) \notag \\
    & \quad + (-\ii)^{\ell+2}\omega x
    [ 1 + (-1)^{\ell+2}\mathcal{R}_{\omega\ell}^P]y_\ell(\omega x)\large\},
\end{align}
where $j_\ell(\rho)$ and $y_\ell(\rho)$ are the spherical Bessel functions of
the first and second kinds, respectively.
Since $y_\ell(\rho)\to\infty$ as $\rho\to 0$, 
for Eqs.~\eqref{eq:u_in_ZFL_asymp_low_omega} and \eqref{eq:u-in-j-and-n} to be compatible, we must have
$[ 1 + (-1)^{\ell+2}\mathcal{R}_{\omega\ell}^P]\to 0$ as $\omega\to 0$ sufficiently fast so that
the second term in Eq.~\eqref{eq:u-in-j-and-n} tends to $0$ in this limit.  Then, since
\begin{equation}
\label{eq:j-ell}
    j_\ell(\rho) \approx \frac{2^\ell \ell!}{(2\ell+1)!}\rho^\ell,    
\end{equation}
\textbf{(}see Ref.~\cite[\S 10.52(i)]{nist}\textbf{)}, we find for 
$ r_h \ll x \ll 1/\omega$,
\begin{equation}
    u_{\omega \ell}^{V,\text{in}} 
    \approx (-\ii)^{\ell+1} \frac{2^{\ell+1}\ell!}{(2\ell+1)!}A_{\omega \ell}^V \omega^{\ell+1}x^{\ell+1}. 
\end{equation}
By comparing this equation and Eq.~\eqref{eq:u_in_ZFL_asymp_low_omega} we find
\begin{equation}
\label{eq:Cin}
    C^{\text{in}}_{\omega\ell} = (-\ii)^{\ell+1}\frac{2^{2\ell+1}(\ell-2)!\ell!(\ell+2)!}{(2\ell)!(2\ell+1)!}
    (M\omega)^\ell \omega A_{\omega\ell}^V,
\end{equation}
for $\omega\approx 0$.

The scalar-type modes  $u^{S, n}_{\omega \ell}$ are obtained from the vector-type modes 
$u^{V, n}_{\omega \ell}$
via the Chandrasekhar transformation~\cite{chandrasekhar_1975,chandrasekhar_1983}. The transformation is 
given, in the $\omega\to 0$ limit, by
\begin{equation}
    \label{eq:chandrasekhar_trasnformation}
        u^{S, n}_{\omega \ell}
=\frac{2}{\sqrt{\Lambda_\ell}}
\left\{
\left[1
+ \frac{2\beta^2}{k_\ell}\,g(r)\right] u^{V, n}_{\omega \ell}
+ \frac{2\beta}{k_\ell}\,\dv{\,u^{V, n}_{\omega \ell}}{x}
\right\}.
\end{equation}
We shall discuss the origin of the normalization factor $2/\sqrt{\Lambda_\ell}$ in the next section.

%%%%%%%%%%%%%%%%%%%%%%%%%
\subsection{Quantization of the gravitational perturbations}
\label{sec:quantization}
%%%%%%%%%%%%%%%%%%%%%%%%%
We first write the Einstein-Hilbert action for pure gravity as
\begin{equation}
    S_{\mathrm{EH}} = \frac{1}{16\pi G}\int R\,\sqrt{-g^{(f)}}\,\dd^4 x, \label{eq:Einstein-Hilbert}
\end{equation}
where $R$ is the Ricci scalar, $G$ is Newton's constant, 
and $g^{(f)}$ denotes the determinant of the full metric $g^{(f)}_{\mu\nu}$. We decompose the full metric as
\begin{equation}\label{eq:g-h-relation}
    g^{(f)}_{\mu\nu} = g_{\mu\nu} + \sqrt{32\pi G}\,h_{\mu\nu},
\end{equation}
where $g_{\mu\nu}$ is the Schwarzschild background metric.
To second order in $h_{\mu\nu}$, the action~\eqref{eq:Einstein-Hilbert} becomes
\begin{equation}
    S_{\textrm{EH}}^{(2)} 
    = \int \mathcal{L}_{\textrm{EH}}^{(2)}\,\sqrt{-g}\, \dd^{4}x,
\end{equation}
where
\begin{eqnarray}
& &\mathcal{L}_{\text{EH}}^{(2)}=
\nabla_\mu h^{\mu\lambda}\nabla^\nu h_{\nu\lambda} 
-\frac{1}{2}\nabla_\lambda h_{\mu\nu}\nabla^\lambda h^{\mu\nu}\notag\\
&& \quad
+\frac{1}{2}\bigl(\nabla^\mu h - 2\nabla_\nu h^{\mu\nu}\bigr)\nabla_\mu h
+ R_{\mu\nu\lambda\sigma}\,h^{\mu\lambda}h^{\nu\sigma},
\end{eqnarray}
with $h\equiv {h^\alpha}_\alpha$. (Note that the tensor indices are raised and lowered by the Schwarzschild background 
metric $g_{\mu\nu}$.) The conjugate momentum current for a solution $h_{\mu\nu}$ is defined as
\begin{eqnarray}
    p^{\alpha\mu\nu} & \coloneqq & \frac{\partial\mathcal{L}^{(2)}_{\textrm{EH}}}{\partial \left(\nabla_\alpha h_{\mu\nu}\right)}.
\end{eqnarray}
Then, the Euler-Lagrange equation is written as
\begin{equation}
    \nabla_\alpha p^{\alpha\mu\nu}  - \frac{\partial \mathcal{L}^{(2)}_{\textrm{EH}}}{\partial h_{\mu\nu}} = 0. \label{eq:EL-canonical}
\end{equation}  
For any two solutions $h^{(i)}_{\mu\nu}$ and $h^{(j)}_{\mu\nu}$ to Eq.~\eqref{eq:EL-canonical}, the following inner product is time-independent~\cite{friedman_1978,wald_2000}:
\begin{equation}
\label{eq:inner_product}
    \big\langle h^{(i)},h^{(j)}\big\rangle \equiv -\ii \int_{\Sigma} \left( \overline{h^{(i)}_{\mu\nu}}p^{\alpha\mu\nu}_{(j)}
    - \overline{p_{(i)}^{\alpha\mu\nu}}h^{(j)}_{\mu\nu}\right)n_\alpha \dd\Sigma,
\end{equation}
where $\Sigma$ is a Cauchy hypersurface with future-pointing unit normal $n^\mu$. The indices $i,j$ denote the set of continuous and discrete labels.

For the perturbation modes presented 
in Sec.~\ref{sec:perturbations}, 
we require the positive-frequency solutions  
to satisfy the normalization 
condition associated with the inner product~\eqref{eq:inner_product}, namely,
\begin{equation}\label{eq:inner-product-for-P}
\big\langle h^{(P,n;\omega \ell m)},\, h^{(P',n';\omega' \ell' m')} \big\rangle
= \delta^{PP'}\delta^{nn'}
  \delta^{\ell\ell'}\delta^{mm'}
  \delta(\omega-\omega'),
\end{equation}
The overall normalization constants $A^{P}_{\omega \ell}$ are determined by combining this normalization condition with the boundary conditions~\eqref{eq:Rin_asymp} and~\eqref{eq:Rup_asymp} as (see Appendix A of Ref.~\cite{bernar_2017}) 
\begin{equation}
    \label{eq:A_V}
    A^{V}_{\omega \ell} = \left[ 8\pi\omega\,(\Lambda_\ell-2)\, \right]^{-1/2},
\end{equation}
and
\begin{equation}
    \label{eq:A_S}
    A^{S}_{\omega \ell} = \left[ 2\pi\omega\,\Lambda_\ell(\Lambda_\ell-2)\, \right]^{-1/2}.
\end{equation}
Then, substituting Eq.~\eqref{eq:A_V} into Eq.~\eqref{eq:Din} we find 
\begin{equation}\label{eq:Cup-solved}
    C_{\omega\ell}^{\text{up}} = -\ii \sqrt{\frac{2\omega}{\pi(\Lambda_\ell-2)}},
\end{equation}
for $\omega\approx 0$.

The normalization factor $2/\sqrt{\Lambda_\ell} = A^S_{\omega\ell}/A^V_{\omega\ell}$ in Eq.~\eqref{eq:chandrasekhar_trasnformation}, which is not in 
Refs.~\cite{chandrasekhar_1975,chandrasekhar_1983}, was needed to make the scalar-type modes defined by Eqs.~\eqref{eq:scalar_ai}--\eqref{eq:scalar_ab} normalized
according to Eq.~\eqref{eq:inner-product-for-P}.

Next, we discuss the quantization of the gravitational perturbations in Schwarzschild spacetime. After fixing the gauge completely, the quantized gravitational field, i.e., the graviton field, is expanded in terms of positive- and negative-frequency modes as
\begin{eqnarray}
\notag \widehat{h}_{\mu\nu}(x)
 = \sum_{P,n} \sum_{\ell,m}
  &&\!\!\!\!\!\! \int_{0}^{\infty} \! \dd\omega \,
   \Big[
      \hat{a}^{P,n}_{\ell m}(\omega)\,
      h^{(P,n;\omega\ell m)}_{\mu\nu}(x)
      \\ &&+ 
      \hat{a}^{P,n\dagger}_{\ell m}(\omega)\,
      \overline{h^{(P,n;\omega\ell m)}_{\mu\nu}(x)}
   \Big],
   \label{eq:graviton_field}
\end{eqnarray}
where $\hat{a}^{P,n}_{\ell m}(\omega)$ and 
$\hat{a}^{P,n\dagger}_{\ell m}(\omega)$ are, respectively, the 
annihilation and creation operators associated with the mode 
$(P,n;\omega\ell m)$. They satisfy the canonical commutation relations
\begin{equation}
\big[\hat{a}^{P,n}_{\ell m}(\omega),\,
      \hat{a}^{P',n'\dagger}_{\ell' m'}(\omega')\big]
= \delta^{PP'}\delta^{nn'}
  \delta^{\ell\ell'}\delta^{mm'}\,
  \delta(\omega-\omega'),
\end{equation}
with all other commutators vanishing.

In the next section, we discuss the interaction of Hawking radiation with the static stress-energy tensor constructed in Sec.~\ref{sec:stress-energy}. In particular, we investigate the corresponding response rate, derived from the one-graviton transition amplitudes.

%%%%%%%%%%%%%%%%%%%%%%%%%
\section{Interaction of Hawking radiation with the static stress-energy tensor}
\label{sec:response-rate}
%%%%%%%%%%%%%%%%%%%%%%%%%

%%%%%%%%%%%%%%%%%%%%%%%%%
\subsection{One-graviton transition amplitudes}
%%%%%%%%%%%%%%%%%%%%%%%%%
The stress-energy tensor associated with a matter action $S_{\textrm{matter}}$ is defined by
\begin{equation}
    T^{\mu\nu} = \frac{2}{\sqrt{-g^{(f)}}}\frac{\delta S_{\textrm{matter}}}{\delta g^{(f)}_{\mu\nu}}.
\end{equation}
For linear coupling between the graviton field and the classical system, using Eq.~\eqref{eq:g-h-relation}, we obtain
\begin{equation}
    \left.\frac{\delta S_{\textrm{matter}}}{\delta h_{\mu\nu}}\right|_{h=0}
    = \sqrt{8\pi G}\, T^{\mu\nu}.
\end{equation}
Hence, the interaction term governing the one-graviton processes induced by 
the classical stress-energy tensor $T^{\mu\nu}$ is
\begin{equation}
\widehat{S}_{\mathrm{int}} = \sqrt{8\pi G}\int T^{\mu \nu}\widehat{h}_{\mu \nu}\sqrt{-g}\,\dd^4x,
\label{eq:int_action}
\end{equation}
where the field operator $\widehat{h}_{\mu \nu}(x)$ is given by Eq.~\eqref{eq:graviton_field}.
We define
\begin{equation}\label{eq:definition-of-gG}
    g_G \coloneqq \sqrt{8\pi G},
\end{equation}
for simplicity. 

The probability amplitudes for the emission and absorption of a graviton 
with $n \in \{\mathrm{in},\mathrm{up}\}$ of the $P$-type, carrying quantum 
numbers $\ell, m$ and frequency $\omega$, in the Boulware~\cite{boulware_1975,boulware_1975b} vacuum state 
$\ket{0_{\mathrm{B}}}$, are given by
\begin{eqnarray}
\mathcal{A}_{\mathrm{em}}^{(P,n;\omega \ell m)} &=& \bra{0_{\mathrm{B}}}\hat{a}^{P,n}_{\ell m}(\omega)  \widehat{S}_{\mathrm{int}} \ket{0_{\mathrm{B}}} \nonumber\\
\label{eq:amplitude_R_emission2}
&=& g_G\int T^{\mu \nu} \overline{h^{(P,n;\omega \ell m)}_{\mu\nu}(x)}\sqrt{-g}\,\dd^4 x, \\
\mathcal{A}_{\mathrm{abs}}^{(P,n;\omega \ell m)} &=& \bra{0_{\mathrm{B}}} \widehat{S}_{\mathrm{int}} \hat{a}^{P,n\,\dagger}_{\ell m}(\omega)\ket{0_{\mathrm{B}}}\nonumber\\
&=& g_G\int T^{\mu \nu} h^{(P,n;\omega \ell m)}_{\mu\nu}(x)\sqrt{-g}\,\dd^4x,
\label{eq:amplitude_R_absorption2}
\end{eqnarray}
respectively. In our case, $T^{\mu \nu}$ is given by Eqs.~\eqref{eq:T_t_t}--\eqref{eq:T_r_r}, and $h^{(P,n;\omega \ell m)}_{\mu\nu}(x)$ is given in Sec.~\ref{sec:perturbations}.

Next, we use the one-graviton transition amplitudes to derive the response rate in the Unruh state.

%%%%%%%%%%%%%%%%%%%%%%%%
\subsection{Response Rate in the Unruh state}
\label{sec:response_rate_Unruh}
%%%%%%%%%%%%%%%%%%%%%%%%
The differential \textit{spontaneous} emission probability, i.e., the 
probability for the emission of  
a single graviton in the mode 
$(P,n;\omega\ell m)$ in the Boulware vacuum state, is
\begin{equation}
d\mathcal{P}^{\mathrm{em}}_{P,n;\omega\ell m}
    = \abs{\mathcal{A}^{(P,n;\omega\ell m)}_{\mathrm{em}}}^{2}
      \, \dd\omega .
\label{eq:probability_Schw_emission}
\end{equation}
The quantum state of a Schwarzschild black hole emitting Hawking radiation~\cite{hawking_1974,hawking_1975} is  
modeled by the Unruh state~\cite{unruh_1976}. In this state, the thermal bath of gravitons in the up-modes is coming from the past event horizon $\mathscr{H}^-$. The expected number of gravitons with frequency $\omega$ in the thermal bath surrounding the source of the stress-energy tensor is
characterized by
\begin{equation}\label{eq:B-E-distribution}
    \langle \hat{a}^{P,\text{up}\,\dagger}_{\ell m}(\omega)\hat{a}^{P',\text{up}}_{\ell' m'}(\omega')\rangle_\upbeta = \delta^{PP'}\delta_{\ell\ell'}\delta_{mm'}n(\omega)\delta(\omega-\omega'),
\end{equation}
where
\begin{equation}
n(\omega) \equiv \frac{1}{\mathrm{e}^{\omega \upbeta}-1},
\label{eq:bath}
\end{equation}
which 
is the Bose-Einstein distribution function at the Hawking 
temperature $\upbeta^{-1} = \kappa/2\pi$, where $\kappa = 1/4M$ is the surface gravity of the Schwarzschild black hole.

If we write the total emission rate for the up-modes as
\begin{align}
\mathcal{R}^{\mathrm{em}}_{\mathrm{tot},\mathrm{up}}
    = \sum_{P=S,V}\sum_{\ell=2}^{\infty}\sum_{m=-\ell}^{\ell}
      \mathcal{R}^{\mathrm{em}}_{P,\mathrm{up};\ell m},
\end{align}
then the emission rate, 
$\mathcal{R}^{\mathrm{em}}_{P,\mathrm{up};\ell m}$, i.e., the emission rate per unit 
time into the mode $(P,\mathrm{up};\ell m)$, is
\begin{align}
\label{eq:emission_rate_Schw}
    \mathcal{R}^{\mathrm{em}}_{P,\mathrm{up};\ell m}
    = \frac{1}{T_0}\int_{0}^{\infty}
      \abs{\mathcal{A}^{(P,\mathrm{up};\omega\ell m)}_{\mathrm{em}}}^{2}
      \,[1+n(\omega)]\, \dd\omega ,
\end{align}
where , $T_0 = \sqrt{f(r_0)}\int_{-\infty}^{\infty} \dd t$ represents the (formally infinite) 
total proper time interval. (The redshift factor $\sqrt{f(r_0)}$ is introduced to convert the coordinate time to the proper time.) 
Equation~\eqref{eq:emission_rate_Schw} follows from 
Eq.~\eqref{eq:probability_Schw_emission} once both spontaneous and induced 
emission processes, the latter arising from the thermal graviton bath  
with the Bose-Einstein distribution function $n(\omega)$, are taken into account.

The absorption rate per unit time 
into the mode $(P,\mathrm{up};\ell m)$ is
\begin{align}
    \mathcal{R}^{\mathrm{abs}}_{P,\mathrm{up};\ell m}
    = \frac{1}{T_0}\int_{0}^{\infty}
      \abs{\mathcal{A}^{(P,\mathrm{up};\omega\ell m)}_{\mathrm{em}}}^{2}
      n(\omega)\, \dd\omega ,
\end{align}
where we have used the relation $
   \mathcal{A}^{(P,n;\omega\ell m)}_{\mathrm{abs}}
    = \overline{\mathcal{A}^{(P,n; \omega\ell m)}_{\mathrm{em}}}$, which follows from Eqs.~\eqref{eq:amplitude_R_emission2} and 
\eqref{eq:amplitude_R_absorption2}, together with 
Eq.~\eqref{eq:probability_Schw_emission} and the expected number of gravitons given by Eq.~\eqref{eq:B-E-distribution}.  
Thus, the \textit{interaction} rate, 
$\mathcal{R}_{P,\mathrm{up};\ell m}$, is the sum of the 
emission and absorption rates,
$\mathcal{R}^{\mathrm{em}}_{P,\mathrm{up};\ell m}$ and 
$\mathcal{R}^{\mathrm{abs}}_{P,\mathrm{up};\ell m}$, respectively. 
Noting that $n(-\omega) = -[1+n(\omega)]$, we obtain  
the general formula for the response rate
for the up-modes as follows:
\begin{align}
\label{eq:diff-interaction-rate-Schw}
    \mathcal{R}_{P,\mathrm{up};\ell m} 
    = \frac{1}{T_0}\int_{-\infty}^{\infty}
      \abs{\mathcal{A}^{(P,\mathrm{up};\omega\ell m)}_{\mathrm{em}}}^{2}
      \,\abs{1+n(\omega)}\, \dd\omega .
\end{align}
 
Now, Eqs.~\eqref{eq:T_t_t}--\eqref{eq:T_r_r}  
show that the only nonzero components of the stress-energy tensor are $T^{tt}$ and $T^{rr}$.
This implies 
that only the scalar-type sector, associated with the gravitational perturbations given by Eqs.~\eqref{eq:scalar_ai}--\eqref{eq:scalar_ab}, contributes to the interaction action. More specifically, only the components $h^{(S,n; \omega \ell m)}_{tt}$ and 
$h^{(S,n; \omega \ell m)}_{rr}$ contribute. (Here, we include the case with 
$n=\mathrm{in}$ because we shall consider
the Hartle-Hawking state later.)  
From Eqs.~\eqref{eq:scalar_ab} and \eqref{eq:Fab_scalar}, we find
$g^{ab}h^{(S,n;\omega\ell m)}_{ab} = 0$, i.e.,
$h^{(S,n; \omega \ell m)}_{rr} = f(r)^{-2}h^{(S,n; \omega \ell m)}_{tt}$. Thus,
\begin{widetext}
    \begin{eqnarray}
        \label{eq:T_h_tt}
        T^{tt} \overline{h_{tt}^{(S,n;\omega \ell m)}} & = &
\frac{\mu}{r^2}
\Big[
\frac{1}{f(r)}\theta(r-r_0) 
+ 
\frac{2}{f'(r)}\delta(r-r_{0}) 
\Big]
\overline{F_{tt}^{(n;\ell)}(t,r)} \overline{Y_{\ell m}(\theta,\phi)}\, \delta^{(2)}_\perp, \\
        \label{eq:T_h_rr}
        T^{rr}\,\overline{h_{rr}^{(S,n;\omega \ell m)}}
& = & -\,\frac{\mu}{r^2f(r)}\,\theta(r-r_{0})\,
   \overline{F_{tt}^{(n;\ell)}(t,r)} \overline{Y_{\ell m}(\theta,\phi)}\,
   \delta^{(2)}_{\perp}.
    \end{eqnarray}
    The function $F_{tt}^{(n;\ell)}$ is obtained from Eq.~\eqref{eq:Fab_scalar}  
    as
  \begin{eqnarray} 
      \label{eq:F_tt_tilde_F_tt}
      F_{tt}^{(n;\ell)}(t,r) &=& \tilde{F}_{tt}^{(n;\ell)}(r) \mathrm{e}^{- \ii \omega t}, \\
    \label{eq:F_tt}
      \tilde{F}_{tt}^{(n;\ell)}(r) & :=& \frac{1}{2H_\ell} \left[f(r)^2 \partial_r^2 - \omega^2\right] \left(r \,H_\ell \, u_{\omega\ell}^{S,n}(r)\right).
  \end{eqnarray}
    
From Eqs.~\eqref{eq:T_h_tt} and \eqref{eq:T_h_rr}, we see that the contribution 
proportional to the Heaviside function $\theta(r-r_{0})$ vanishes in $T^{\mu \nu}\,h_{\mu \nu}^{(S,n;\omega \ell m)}$ and, consequently, in the one-graviton transition amplitudes. (This may be understood as the cancellation  
of the explicit string contribution under  
the standard choice of gauge, which we have adopted here. 
As in the Rindler case~\cite{brito_2024_gw}, this is a 
gauge-dependent feature:  
A different gauge choice can lead to a nonzero string contribution, 
although the final physical result for 
the emission and absorption rates will remain gauge invariant.)
Hence, the emission amplitude $\mathcal{A}^{(S,n;\omega \ell m)}_{\mathrm{em}}$ in terms of $F_{tt}^{(n;\ell)}(t,r)$ is given by
\begin{eqnarray}
    \label{eq:emission_amplitude_expli_1}
   \mathcal{A}^{(S,n;\omega \ell m)}_{\mathrm{em}} &=& 2 g_G\mu
\int \dd^4x \,\delta(r-r_{0}) \,\delta(\theta-\theta_{0}) \,\delta(\phi-\phi_{0}) \,
\,f'(r)^{-1}\,
\overline{F_{tt}^{(n;\ell)}(t,r)} \overline{Y_{\ell m}(\theta,\phi)}\, \delta^{(2)}_\perp, \\
\label{eq:emission_amplitude_expli_2}
&=& 2 g_G\mu f'(r_0)^{-1}\,I(\omega)\,\overline{\tilde{F}_{tt}^{(n;\ell)}(r_0)} \,\overline{Y_{\ell m}(\theta_0,\phi_0)}, \\
\label{eq:emission_amplitude_expli_3}
&=& g_Gm_0\,f(r_0)^{-1/2}\,I(\omega)\,\overline{\tilde{F}_{tt}^{(n;\ell)}(r_0)} \,\overline{Y_{\ell m}(\theta_0,\phi_0)}.
\end{eqnarray}
Here, we have defined 
$I(\omega)\coloneqq 2\pi \delta(\omega)$ and, in the last equality, we have used Eq.~\eqref{eq:mu}. Recall that, since the stress-energy tensor is static, it interacts only with zero-frequency gravitons of the thermal bath.\footnote{The distinction between the emission and absorption becomes blurred
for $\omega=0$. If we considered our system of a point mass and a string as the static limit of a non-static system, as was done in Refs~\cite{crispino_1998,higuchi_1997,higuchi_1998}, then the response rate would be the sum of the emission and absorption rates.} Hence, we are interested in the low-frequency regime of the graviton modes obtained in Sec.~\ref{sec:gw_ZFL_modes}.  
We find from Eqs.~\eqref{eq:u_up_ZFL_sol} and \eqref{eq:Cup-solved} that  
the  
radial function $u_{\omega\ell}^{S,\text{up}}$ behave  
like $\sqrt{\omega}$ in the low-frequency regime. Let us define the $\omega$-independent function $\tilde{u}_{\ell}^{S,\text{up}}(r)$ through
\begin{equation}
    \label{eq:uS_tilde}
    \tilde{u}_{\ell}^{S,\text{up}} \coloneqq \lim_{\omega\to 0} u_{\omega\ell}^{S,\text{up}}/\sqrt{\omega}.
\end{equation}
Then, from Eqs.~\eqref{eq:F_tt_tilde_F_tt}--\eqref{eq:emission_amplitude_expli_3}, we find 
\begin{equation}
    \label{eq:emission_amplitude_expli}
   \mathcal{A}^{(S,\text{up};\omega \ell m)}_{\mathrm{em}} = g_Gm_0\,I(\omega)\,\left\{\frac{f(r_0)^{3/2}}{2H_\ell} \left[\overline{\dv[2]{\,}{r_0} \left(r_0 \,H_\ell \, \tilde{u}_{\ell}^{S,\text{up}}(r_0)\right)}\right]\sqrt{\omega} + \mathcal{O}\left(\omega^{3/2}\right) \right\}\,\overline{Y_{\ell m}(\theta_0,\phi_0)}.
\end{equation}
\end{widetext} 
This equation shows that $\mathcal{A}^{(S,\text{up};\omega \ell m)}_{\mathrm{em}} \to 0$ as $\omega \to 0$. 
Thus, the response rate vanishes when the initial state is the Boulware state.
This is because the stress-energy tensor describes a static system, and the spontaneous emission rate vanishes. However, the induced emission rate remains nonzero due to the 
interaction of the source with zero-energy gravitons from the thermal bath of Hawking radiation, whose expected number of gravitons diverges in the infrared limit. Hence, the presence of Hawking radiation is crucial for  
a nonvanishing response rate. 

For low frequencies, we approximate the Bose-Einstein  
distribution function as $n(\omega) \approx 1/\upbeta \omega$, where $\upbeta^{-1} = \kappa/2\pi = (8\pi M)^{-1}$. By substituting 
Eqs.~\eqref{eq:emission_amplitude_expli} into  Eq.~\eqref{eq:diff-interaction-rate-Schw}, we obtain
\begin{eqnarray}
        \mathcal{R}_{\ell m} 
    & = & \frac{g_G^2 m_0^2}{4M}\,f(r_0)^{\frac{5}{2}} \left[\frac{1}{2H_\ell}\dv[2]{\,}{r_0}\!\!\left(r_0 H_\ell \,\tilde{u}_{\ell}^{S,\text{up}} 
    \right)\right]^2 \nonumber \\  
    && \times\abs{Y_{\ell m}(\theta_0,\phi_0)}^2, \label{eq:response_rate1}
\end{eqnarray}
where we have used the standard formal expression $\abs{I(\omega)}^2 =  2\pi  T_0 \delta(\omega)/\sqrt{f(r_0)}$. Note that only the scalar-type up-modes ($P = S$ and $n = \text{up}$) yield a nonvanishing contribution to the response rate~\eqref{eq:response_rate1} in the Unruh state. 
We can sum over the azimuthal quantum number $m$ using
\begin{equation}
\sum_{m=-\ell}^{\ell} \abs{Y_{\ell m}(\theta_0,\phi_0)}^{2}
= \frac{2\ell+1}{4\pi},
\end{equation}
as
\begin{eqnarray}
        \mathcal{R}_{\ell} & \coloneq &\!\! \sum_{m=-\ell}^\ell \mathcal{R}_{\ell m} \nonumber \\
    & = &\frac{(2\ell+1)g_G^2 m_0^2}{16\pi M}\! \nonumber\\
    & &\times f(r_0)^{\frac{5}{2}}\! 
    \left[\frac{1}{2H_\ell}\dv[2]{\,}{r_0}\!\!\left(r_0 H_\ell \,\tilde{u}_{\ell}^{S,\text{up}} \right)\right]^2.
     \label{eq:response_rate}
\end{eqnarray}
This equation is 
the response rate associated with a fixed value of the angular momentum quantum number $\ell$. The expression in square brackets is  
given in terms of the Legendre function as 
\begin{equation}
    \label{eq:expre_appendix_A}
    \frac{1}{2H_\ell}\dv[2]{\,}{r_0}\left(r_0 H_\ell \,\tilde{u}_{\ell}^{S,\text{up}} \right) =-\ii \sqrt{\frac{2}{\pi\,k_\ell}}\,r_0^2 \, \dv[2]{}{r_0}Q_\ell \left(\frac{r_0}{M}-1 \right).
\end{equation}
\textbf{(}See Appendix~\ref{sec:appendix_A} for a derivation of this formula.\textbf{)}
The total response rate   
is then 
\begin{eqnarray}
    \label{eq:Tot_response_rate}
        \mathcal{R}_{\text{tot}} 
    & = & \frac{g_G^2 m_0^2}{8\pi^2 M}\,r_0^4\,f(r_0)^{\frac{5}{2}}\nonumber \\ 
    & & \times \sum_{\ell \geq 2}\frac{(2\ell +1)}{\Lambda_\ell(\Lambda_\ell-2)}\left[\dv[2]{}{r_0}Q_\ell \left(\frac{r_0}{M}-1 \right)\right]^2.
\end{eqnarray}
The summation over $\ell$ can be performed using the $n=2$ case of the formula derived in Appendix~\ref{sec:appendix_B}. 
Thus, we obtain for the total response rate, with respect to the proper time at the point mass, of the stress-energy tensor given by Eqs.~\eqref{eq:T_h_tt}--\eqref{eq:T_h_rr} in the Unruh state:
\begin{equation}
    \label{eq:Tot_response_rate_final}
        \mathcal{R}_{\text{tot}} 
    = \frac{Gm_0^2}{\pi\hbar c^2}\,\varrho(r_0)\,\left[-\frac{2r_h}{3}\dv{\,\,}{r_0}Q_2 \left(\frac{2r_0}{r_h}-1 \right)\right],
\end{equation}
where, we have restored the constants $G$, $c$ and $\hbar$ using
Eq.~\eqref{eq:definition-of-gG} and dimensional analysis, and where $\varrho(r_0)$ is the proper acceleration at $r=r_0$ given by Eq.~\eqref{eq:acceleration}, i.e.,
\begin{equation}
    \varrho(r_0) = \frac{G M}{r_0^2}\left( 1- \frac{r_h}{r_0}\right)^{-1/2}.
\end{equation} 
(Note that $r_h = 2 G M/c^2$.)
Equation~\eqref{eq:Tot_response_rate_final} shows that the total response rate in the Schwarzschild spacetime is finite for $r_0 > r_h$, as in the electromagnetic~\cite{crispino_1998} and scalar~\cite{higuchi_1997,higuchi_1998} cases. 
This contrasts with the Rindler case, in which the total response rate exhibits 
a power-law infrared divergence~\cite{brito_2024_gw}. Hence, in the limit $r_0\to r_h$ the total response rate $\mathcal{R}_{\text{tot}}$ is expected to agree
with the result of Ref.~\cite{brito_2024_gw} in Rindler spacetime with an infrared cutoff because the stress-energy
tensor becomes that in Rindler spacetime studied in Ref.~\cite{brito_2024_gw} in this limit \textbf{(}see Eqs.~\eqref{eq:T_tt_Rindler} and \eqref{eq:T_rr_Rindler}\textbf{)}. 
To see that this is indeed the case, note that in the near-horizon region, $r_0 \gtrsim r_h$, the quantity inside the 
square brackets in Eq.~\eqref{eq:Tot_response_rate_final} can be approximated using
\begin{equation}
    \label{eq:Q2_approx}
    -\dv{\,\,}{r_0}Q_2 \left(\frac{2r_0}{r_h}-1 \right) \approx \frac{2r_h}{c^4}\,\varrho(r_0)^{2}, \quad \text{for\,\,} r_0 \gtrsim r_h.
\end{equation}
 Thus, the total response rate \eqref{eq:Tot_response_rate_final} in the near horizon region is given by
\begin{equation}
\label{eq:Tot_response_rate_near_horizon}
        \mathcal{R}_{\text{tot}} 
    \approx \frac{Gm_0^2}{\pi\hbar c^2}\,\varrho(r_0)\,\left[\frac{4r_h^2}{3c^4}\,\varrho(r_0)^2\right],\quad \text{for\,\,} r_0 \gtrsim r_h.
\end{equation}

In the Rindler case, the total response rate, 
$\mathcal{R}_{\mathrm{tot}}^{\mathrm{Rindler}}$, can be obtained from 
Eq.~(62) of Ref.~\cite{brito_2024_gw} by integrating the differential response 
rate (with the constants $G$, $c$ and $\hbar$ restored),
\begin{equation}
    \mathcal{R}_{\mathbf{k}_\perp} = \frac{Gm_0^2c^2}{\pi^2\hbar a}\left|K_2\left(\frac{k_\perp c^2}{a}\right)\right|^2,
\end{equation} 
where $a$ is the proper acceleration of the point mass,
over the transverse momentum, $\mathbf{k}_\perp$, with $k_\perp\coloneqq \lVert\mathbf{k}_\perp\rVert$. 
This quantity exhibits an infrared divergence.  
We introduce an infrared cutoff $k_\perp^0$ and integrate only over $\mathbf{k}_\perp$ satisfying $k_\perp \geq k_\perp^0$. 
One then finds
\begin{eqnarray}
\mathcal{R}_{\mathrm{tot}}^{\mathrm{Rindler}}
\label{eq:Total_response_rindler_k}
&=& 2\pi \int_{k_\perp^0}^{\infty} \mathrm{d}k_\perp\, k_\perp\,
    \mathcal{R}_{\mathbf{k}_\perp} \\
\label{eq:Total_response_rindler_near_horizon}
&\approx& \frac{4Gm_0^{2}a^3}{\pi\hbar c^6(k_\perp^{0})^{2}},
\end{eqnarray}
to leading order in $1/k_\perp^0$.  This result agrees with Eq.~\eqref{eq:Tot_response_rate_near_horizon} if we let
$k_\perp^0 = \sqrt{3}/r_h$.  Thus, the near-horizon limit of the response rate for a static point mass in Schwarzschild spacetime 
is reproduced from that for the uniformly accelerated point mass in Minkowski spacetime with the cutoff for 
the graviton transverse momentum of the order of the
inverse Schwarzschild radius.

%%%%%%%%%%%%%%%%%%%
\subsection{Response Rate in the Hartle-Hawking state}
\label{sec:hartle-hawking}
%%%%%%%%%%%%%%%%%%%
The Hartle-Hawking state~\cite{hartle_1976} is characterized by a thermal flux at 
the Hawking temperature incoming from past null infinity, $\mathscr{I}^-$, in 
addition to the thermal flux coming from the past event horizon, 
$\mathscr{H}^-$. In this state, the black hole is in thermal equilibrium in the 
sense that there is no net flux of particles at infinity. The contribution to the 
response rate arising from the thermal flux emanating from  
$\mathscr{H}^-$ was 
computed in Sec.~\ref{sec:response_rate_Unruh}. Here, we  
show that the contribution from modes incoming from 
$\mathscr{I}^{-}$ to the response rate of the stress-energy tensor,  
given by Eqs.~\eqref{eq:T_t_t} and~\eqref{eq:T_r_r}, vanishes. 

The total response rate in the Hartle-Hawking state, $\mathcal{R}^{\text{HH}}_{\mathrm{tot}},$ is given by
\begin{align}
\label{eq:response_rate_hartle_hawking}
\mathcal{R}^{\text{HH}}_{\mathrm{tot}}
    = \sum_{P=S,V}\sum_{\ell=2}^{\infty}\sum_{m=-\ell}^{\ell}
      \left[\mathcal{R}_{P,\mathrm{up};\ell m}+\mathcal{R}_{P,\mathrm{in};\ell m}\right],
\end{align}
where the response rate, 
$\mathcal{R}_{P,n;\ell m}$, is the sum of the
emission and absorption rates,
$\mathcal{R}^{\mathrm{em}}_{P,n;\ell m}$ and $ 
\mathcal{R}^{\mathrm{abs}}_{P,n;\ell m}$, respectively. The response rate $\mathcal{R}_{P,\mathrm{up};\ell m}$ have been computed in Sec.~\ref{sec:response_rate_Unruh}.
To compute $\mathcal{R}_{P,\mathrm{in};\ell m}$ first note that 
\begin{equation}
\label{eq:diff-interaction-rate-Schw_hartle-hawking}
    \mathcal{R}_{P,\mathrm{in};\ell m} 
    = \frac{1}{T_0}\int_{-\infty}^{\infty}
      \abs{\mathcal{A}^{(P,\mathrm{in};\omega\ell m)}_{\mathrm{em}}}^{2}
      \,\abs{1+n(\omega)}\, \dd\omega,
\end{equation}
which can be found by the same argument that led to Eq.~\eqref{eq:diff-interaction-rate-Schw} for the up-modes.
Similarly to the up-mode case, the emission amplitude can be obtained from Eqs.~\eq{eq:F_tt} and~\eq{eq:emission_amplitude_expli_3}, and the low-frequency in-mode solutions, $u_{\omega\ell}^{S,\text{in}},$ are obtained from 
Eqs.~\eqref{eq:u_in_ZFL_sol}, \eqref{eq:Cin}, \eqref{eq:chandrasekhar_trasnformation}, and 
\eqref{eq:A_V}.  
Now, Eqs.~\eqref{eq:Cin} and 
\eqref{eq:A_V} imply that  
the functions $u_{\omega\ell}^{S,\text{in}}$ behave  
like $\omega^{\ell+1/2}$ in the low-frequency regime. Consequently, the corresponding
(spontaneous) emission 
probability density, $|\mathcal{A}^{(S,\mathrm{in};\omega\ell m)}_{\mathrm{em}}|^{2}$, scales as $\omega^{2\ell+1}$, whereas the thermal factor 
is approximated by $n(\omega) \approx 1/(\upbeta\omega)$ for small $\omega$. It follows 
that the contribution of the in-modes to the response rate in the Hartle-Hawking state vanishes, i.e., $\mathcal{R}_{P,\mathrm{in};\ell m} = 0,$
because $\ell\geq 2$. This implies
\begin{equation}
    \label{eq:total_response_Hartle-Hawking}
    \mathcal{R}^{\text{HH}}_{\mathrm{tot}} = \mathcal{R}_{\mathrm{tot}},
\end{equation}
where $\mathcal{R}_{\mathrm{tot}}$ is given by Eq.~\eqref{eq:Tot_response_rate_final}.
Therefore, the total response rate of the stress-energy tensor, given by Eqs.~\eqref{eq:T_t_t}--\eqref{eq:T_r_r}, in the Hartle-Hawking state is 
identical to that obtained in the Unruh state.

This equality also holds 
for the electromagnetic field.
We briefly discuss it here since this observation was not made in  
Ref.~\cite{crispino_1998}, which gave the response rate of a static electric charge
in the Unruh state.  
We first note that the in-mode for the electromagnetic field,
which is relevant to
the response rate to the flux coming from $\mathscr{I}^-$ of a static electric charge $q$ at $r=r_0$, described by the current $j^{\mu} = (j^t,0,0,0)$, with $j^t\! =\! q\,\delta(r-r_0)\,\delta_\perp^{(2)}\!/r^2$, is given by\footnote{The  
modes $A_\mu^{(\text{in},\omega\ell m)}$ can be obtained from 
Ref.~\cite[Eqs. (3.5) and (3.6)]{crispino_2001}.}
\begin{equation}
    \label{eq:elet_field}
    A_\mu^{(\text{in},\omega\ell m)} \! =\! B_{\omega \ell} \left(\! -\ii f(r) \dv{\varphi^{\text{in}}_{\omega \ell}}{r} Y_{\ell m}, \frac{\omega\varphi^{\text{in}}_{\omega \ell}}{f(r)}\,Y_{\ell m}, 0, 0 \right) \mathrm{e}^{- \ii \omega t},
\end{equation}
where $\ell\geq 1$ and 
\begin{equation} 
B_{\omega \ell} = \left(4\pi\omega \Lambda_\ell \right)^{-1/2}.
\end{equation} 
For low frequencies one finds
\begin{equation}
    \label{eq:photon_ZFL}
    \varphi^{\text{in}}_{\omega \ell} \approx D^{\text{in}}_{\omega \ell}\, r\,F_\ell, 
\end{equation}
where $D^{\text{in}}_{\omega \ell}$ is a normalization constant and where $F_\ell$ is the solution of Eq.~\eqref{eq:Z-equation} with $\rho_{s=1} = 2,$ which  
does not diverge as $r\to r_h$, i.e., 
\begin{equation}
    \label{eq:F_photon}
    F_\ell(z) = P_\ell(z) - \frac{z-1}{\Lambda_\ell}\,\dv{\,}{z}P_\ell(z).
\end{equation}
Recall that $z=r/M-1.$ 

On the one hand, using Eq.~\eqref{eq:legendreP_asymp}, we find that, for large $r$,
\begin{equation}
    \label{eq:in_asymp}
    \varphi^{\text{in}}_{\omega \ell} \approx D^{\text{in}}_{\omega \ell} \,\frac{(2\ell)!}{2^\ell (\ell-1)!(\ell+1)!}\frac{r^{\ell+1}}{M^\ell}.
\end{equation}
On the other hand, the in-mode  
radial function $\varphi^{\text{in}}_{\omega \ell}$ can also be written in this regime as
\begin{equation}
    \label{eq:in_mode_j_ell}
    \varphi^{\text{in}}_{\omega \ell} \approx 2 B_{\omega \ell}\,\omega x\,j_{\ell}(\omega x), \quad \text{for\,\,} x \gg r_h,
\end{equation}
where $j_\ell$ 
is the spherical Bessel function of the first kind
and where the tortoise coordinate $x$ is given by Eq.~\eqref{eq:tortoise}. 
Thus, using Eq.~\eqref{eq:j-ell} for $r_h\ll x \ll 1/\omega$  
we find
\begin{equation}
    \label{eq:in_asymp_photon}
    \varphi^{\text{in}}_{\omega \ell} \approx \frac{2^{\ell+1} \ell!}{(2\ell+1)!} B_{\omega \ell}\, (\omega x)^{\ell+1}.
\end{equation} 
By comparing Eqs.~\eq{eq:in_asymp} and \eq{eq:in_asymp_photon}, we find 
\begin{equation}
    \label{eq:Din_photon}
    D^{\text{in}}_{\omega \ell} = \frac{2^{2\ell+1}(\ell-1)!\,\ell!\,(\ell+1)!}{(2\ell+1)!\,(2\ell)!}\,M^\ell\omega^{\ell+1} B_{\omega \ell}.
\end{equation}
Then, substituting this formula 
into Eq.~\eqref{eq:photon_ZFL}, we find that 
the radial function for the in-modes of the photon, $\varphi^{\text{in}}_{\omega \ell}$, are given as
\begin{equation}
    \label{eq:in_mode_photon_final}
    \varphi^{\text{in}}_{\omega \ell} \approx \frac{2^{2\ell+1}(\ell-1)!\,\ell!\,(\ell+1)!}{\sqrt{4\pi\,\Lambda_\ell}\,(2\ell+1)!\,(2\ell)!}\,M^\ell\omega^{\ell+1/2} r\,F(r),
\end{equation}
for $\omega \approx 0.$
This equation implies that the electromagnetic 
in-modes also behave like $\omega^{\ell+1/2}$ in the low-frequency regime.
Then, the response rate of the static charge to the flux of thermal bath coming from $\mathscr{I}^{-}$ can be shown to
vanish as in the graviton case since $\ell\geq 1$. 
Thus, the static electric charge also responds in the Unruh and Hartle-Hawking states in exactly the same way.

The gravitational and electromagnetic cases contrast with the scalar case,  
which have been  
studied in Refs.~\cite{higuchi_1997,higuchi_1998}.
The response rate of a point scalar source at $r=r_0$ in the Unruh and Hartle-Hawking states are not equal, unless the scalar field is taken to be the massless limit of the massive scalar field in Schwarzschild spacetime. One can see from Refs.~\cite{higuchi_1997,higuchi_1998} that the radial scalar modes also scales as 
 $\omega^{\ell+1/2}$ in the low-frequency regime; however, in this case the $\ell=0$ mode gives a nonzero response rate to the flux coming from $\mathscr{H}^-$. Thus, only the scalar field, which has a physical monopole mode, yields a nonvanishing contribution of the in-modes to the response rate.

For convenience, Appendix~\ref{sec:appendix_D} compiles the response rates of point particle systems interacting with bosonic fields, including both previously derived results and those obtained in the present work, in some spacetime backgrounds and quantum states.

\section{Final Remarks}
\label{sec:remarks}
%%%%%%%%%%%%%%%%%%%%%%%%%%%%%%%%%
We investigated the interaction of the gravitational Hawking radiation with a static, 
conserved stress-energy tensor outside a Schwarzschild black 
hole
with the graviton field in the Unruh state. 
We also considered the Hartle-Hawking state, which models a black hole in thermal equilibrium with a thermal bath at the Hawking temperature.
The stress-energy tensor describes a 
point mass held at a fixed radial position $r_0$ by a one-dimensional radial 
string extending from $r_0$ to infinity and
satisfies the weak energy condition.  
In particular, we  
calculated the response rate of  
this stress-energy tensor to the thermal bath of gravitons.

We found that the total response rate of the stress-energy tensor to the gravitational 
Hawking radiation from the Schwarzschild black hole, i.e., in the Unruh state, is finite for $r_0>r_h$.
We found this response rate in closed form as had been done 
for the scalar and electromagnetic cases~\cite{crispino_1998,higuchi_1997,higuchi_1998}. 
The finiteness of this response rate  
contrasts with the analogous case in Rindler spacetime, where the total response rate exhibits an infrared power-law divergence~\cite{brito_2024_gw}.
This suggests that,  
as $r_0 \to r_h$, the response rate in Schwarzschild spacetime approaches that in Rindler spacetime with an infrared cutoff.
We found that this is indeed the case.  
Comparing the total response rate near the black hole horizon with that in the Rindler case, we showed that the size of the black hole acts as a natural cutoff for the infrared divergence.

We also found that the stress-energy tensor does not respond to the thermal flux of gravitons incoming from the past event 
horizon in the Hartle-Hawking state. That is, its total response rate is the same in the Unruh and Hartle-Hawking states.  
This contrasts with the corresponding  
results for the scalar field~\cite{higuchi_1997,higuchi_1998} but  
is the same as that in the case with  
the electromagnetic field~\cite{crispino_1998}. 
In the scalar case, the in-mode  with angular momentum $\ell=0$ gives a nonvanishing contribution to the response rate in the Hartle-Hawking state.  In all three cases 
the square of the in-modes contributing to the response rate, i.e., the ones that couple to the classical source 
associated with the point particle, behave  
like  $\omega^{2\ell+1}$ 
in the low-frequency regime, while the thermal factor $n(\omega)$ behaves 
like $\omega^{-1}$ for low frequencies.
Thus, only the scalar field, which has a physical monopole, i.e., the mode with $\ell=0$, yields a nonvanishing response for the point-particle source.

It is important to recall that the total response rate computed here corresponds to the number of interactions between the point particle and the thermal bath associated with the quantum state per unit proper time. Specifically, it gives the rate of absorption and stimulated emission events experienced by the classical system per unit proper time of an observer at $r=r_0$. For a point particle identified with an electron (proton) held static at the photon sphere of a solar-mass black hole, the characteristic proper time between interactions with a graviton from the Hawking radiation is of order $10^{34}$ years ($10^{27}$ years), whereas the corresponding interaction time with a photon is of order $10^{-2}$ seconds. Nevertheless, since the graviton response rate grows faster than the electromagnetic one as the horizon is approached, interactions with gravitons become dominant sufficiently close to the black hole.

%%%%%%%%%%%%%%%%%%%%%%%%%%%%%%
\begin{acknowledgments}
%%%%%%%%%%%%%%%%%%%%%%%%%%%%%%
The authors thank %Funda\c{c}\~ao Amaz\^onia de Amparo a Estudos e Pesquisas (FAPESPA),  
Conselho Nacional de Desenvolvimento Cient\'ifico e Tecnol\'ogico (CNPq) and Coordena\c{c}\~ao de Aperfei\c{c}oamento de Pessoal de N\'{\i}vel Superior (Capes) - Finance Code 001, in Brazil, for partial financial support.
JB thanks the University of York in England for the kind hospitality.
%during the development and completion of this work.
This work has further been supported by the 
%European Union's Horizon 2020 research and innovation (RISE) programme H2020-MSCA-RISE-2017 Grant No. FunFiCO-777740 and by the 
European Horizon Europe staff exchange (SE) programme HORIZON-MSCA-2021-SE-01 Grant No. NewFunFiCO-101086251.
\end{acknowledgments}

\begin{widetext}
%%%%%%%%%%%%%%%%%%%%%%%%%%%%
\appendix
%%%%%%%%%%%%%%%%%%%%%%%%%%%%%
%%%%%%%%%%%%%%%%%%%%%%%%%%%%
\section{Derivation of Eq.~\eqref{eq:expre_appendix_A}}
\label{sec:appendix_A}
%%%%%%%%%%%%%%%%%%%%%%%%%%%%
In this appendix, we derive the low-frequency expressions for the radial functions for the scalar-type up-modes, 
$u_{\omega\ell}^{S,\mathrm{up}}$, in terms of the Legendre functions of the second 
kind, $Q_\ell(z)$, which 
are obtained from the corresponding radial functions for the vector-type modes,
$u_{\omega\ell}^{V,\mathrm{up}}$,  
through the Chandrasekhar 
transformations~\cite{chandrasekhar_1975,chandrasekhar_1983}. 
We then derive Eq.~\eqref{eq:expre_appendix_A}, which gives the $\omega\to 0$ limit of the radial functions for the components $h_{tt}^{(S,\text{up},\omega\ell m)}$ divided by $\sqrt{\omega}$.  The expressions for 
the radial functions for the in-modes, $u_{\omega\ell}^{S,\mathrm{in}}$, can be derived in  
a similar manner. 

Let us define [see Eq.~\eqref{eq:ZFL_regge-wheeler_sol}]
\begin{equation}\label{eq:Z-tilde-def}
    \widetilde{Z}_\ell(z) \coloneqq \left[ 1 - 4\frac{(z-1)}{\Lambda_\ell}\frac{\mathrm{d}\ }{\mathrm{d}z}
    + 6\frac{(z-1)^2}{k_\ell}\frac{\mathrm{d}^2\ }{\mathrm{d}z^2}\right]Q_\ell(z).
\end{equation}
Then, we find from Eq.~\eqref{eq:u_up_ZFL_sol}
\begin{equation}
    u^{V,\text{up}}_{\omega\ell} = C^{\text{up}}_{\omega\ell}M(z+1)\widetilde{Z}_\ell(z),
\end{equation}
for small $\omega$.
We use the Legendre equation satisfied by $Q_\ell(z)$ in Eq.~\eqref{eq:Z-tilde-def} to express $Q_\ell''(z)$ in terms of
$Q_\ell(z)$ and $Q_{\ell}'(z)$ to obtain
\begin{align}\label{eq:Z-tilde-explicit}
    \widetilde{Z}_\ell(z) =\left[ 1 + \frac{6(z-1)}{(\Lambda_\ell-2)(z+1)}\right]Q_\ell(z)
    - 4(z-1)\left[ \frac{1}{\Lambda_\ell} + \frac{3z}{k_\ell(z+1)}\right]Q_\ell'(z).
\end{align}
We again use the Legendre equation in this manner and obtain
\begin{align}\label{eq:Z-tilde-diff-explicit}
    \widetilde{Z}_\ell'(z)
    & =  -4\left[\frac{1}{z+1} + \frac{3(z-1)}{(\Lambda_\ell-2)(z+1)^2}\right]Q_\ell(z)
    + \left[ 1+\left(\frac{4}{\Lambda_\ell}+ \frac{6}{\Lambda_\ell-2}\right)\frac{z-1}{z+1}
    + \frac{12}{k_\ell}\left(\frac{z-1}{z+1}\right)^2\right]Q_\ell'(z).
\end{align}

We define
\begin{equation}
 Y_\ell(z)\coloneqq  \frac{rH_\ell}{M^2}\left\{  \left[ 1 + \frac{72M^2}{k_\ell} \frac{f(r)}{r^2H_\ell}\right] r\widetilde{Z}_\ell(z) 
+ \frac{12M}{k_\ell}f(r)\frac{\mathrm{d}\ }{\mathrm{d}r}[r\widetilde{Z}_\ell(z)]\right\} .
\end{equation}
Then, we find from Eq.~\eqref{eq:chandrasekhar_trasnformation}
\begin{equation}
    u^{S,\text{up}}_{\omega\ell} = \frac{2}{\sqrt{\Lambda_\ell}}\frac{C_{\omega\ell}^{\text{up}}M^2}{rH_\ell}Y_\ell(z),
\end{equation}
for small $\omega$.
Then, the function $\tilde{u}^{S,\text{up}}_{\ell}$ defined by Eq.~\eqref{eq:uS_tilde} is given by
\begin{equation}\label{eq:uS_up-final}
    \tilde{u}^{S,\text{up}}_{\ell} = -\ii\sqrt{\frac{8}{\pi k_\ell}}\,\frac{M^2}{rH_\ell}Y_\ell(z).
\end{equation}
Now,
\begin{eqnarray}
 Y_\ell(z) & = & \left[ (\Lambda_\ell-2)(z+1)^2+6(z+1) + \frac{144(z-1)}{k_\ell(z+1)}
+ \frac{12}{\Lambda_\ell}(z-1)\right]\widetilde{Z}_\ell(z) +12(z-1)\left[\frac{1}{\Lambda_\ell}(z+1)+\frac{6}{k_\ell}\right]\widetilde{Z}_\ell'(z).
\end{eqnarray}
Substituting Eqs.~\eqref{eq:Z-tilde-explicit} and \eqref{eq:Z-tilde-diff-explicit} into this equation, we find
\begin{equation}
    Y_\ell(z) = (z+1)\left[ (\Lambda_\ell+4)z+\Lambda_\ell-2\right]Q_\ell(z)
    -\frac{4(\Lambda_\ell+1)}{\Lambda_\ell}(z^2-1)(z+1)Q_\ell'(z).
\end{equation}
We differentiate this expression twice and use
\begin{eqnarray}
    Q_\ell(z) & = & \frac{1}{\Lambda_\ell}\left[ (z^2-1)Q_\ell''(z) + 2zQ_\ell(z)\right],\\
  (z^2-1)Q_\ell'''(z) & = & -4zQ_\ell''(z) + (\Lambda_\ell-2)Q_\ell'(z),  
\end{eqnarray}
to find
\begin{eqnarray}
    Y_\ell''(z) & = & (z+1)[(\Lambda_\ell-2)(z+1) + 6] Q_\ell''(z) \nonumber \\
    & = & \frac{r^2}{M^2}H_\ell Q_\ell''(z).
\end{eqnarray}
Using this equation and Eq.~\eqref{eq:uS_up-final}, we find Eq.~\eqref{eq:expre_appendix_A}.

%%%%%%%%%%%%%%%%%%%%%%%%%%%%
\section{A summation formula for the Legendre functions of the second kind}
\label{sec:appendix_B}
%%%%%%%%%%%%%%%%%%%%%%%%%%%%

Define for $n\geq 1$
\begin{equation}
S_n(z) = \sum_{\ell=n}^\infty
\frac{2\ell+1}{\Lambda_\ell(\Lambda_\ell-2)\cdots[\Lambda_\ell-n(n-1)]}\left[
\dv[n]{\,}{z}Q_\ell(z)\right]^2,
\end{equation}
where $\Lambda_\ell=\ell(\ell+1)$ as before. We can show that this series
is absolutely convergent if $z > 1$ by using the following 
formula derived from Ref.~\cite[Eq.~8.821.3]{gradshteyn}:
\begin{equation}
    \dv[n]{\,}{z}Q_\ell(z) = \frac{(-1)^n(\ell+n)!}{2^{\ell+1}\ell !}
    \int_{-1}^1 (1-t^2)^\ell (z-t)^{-\ell-n-1}\mathrm{d}t,
\end{equation}
and noting that the maximum value of $(1-t^2)/(z-t)$ for $-1<t < 1$ is
$2(z-\sqrt{z^2-1})$ at $t=z-\sqrt{z^2-1}$.  Thus,
\begin{eqnarray}
    \left| \dv[n]{\,}{z}Q_\ell(z)\right| \leq
    \frac{(\ell+n)!}{2\cdot \ell !}
    (z+\sqrt{z^2-1})^{-\ell}\int_{-1}^1\frac{\mathrm{d}t}{(z-t)^{n+1}}.
\end{eqnarray}
Then,
\begin{equation}
    \limsup_{\ell\to\infty}\left|\dv[n]{\,}{z}Q_{\ell+1}(z)\Big/
    \dv[n]{\,}{z}Q_\ell(z)\right|^2 \leq \frac{1}{(z+\sqrt{z^2-1})^2} < 1.
\end{equation}
Hence, the series $S_n(z)$ is absolutely convergent if $z>1$.
In this appendix, we prove that
\begin{equation}
    S_n(z) = (-1)^{n+1}\frac{2n}{(2n-1)!!}\frac{1}{(z^2-1)^{n+1}}\dv[n-1]{\,}{z}Q_n(z).  \label{eq:summation-formula}
\end{equation}
The case $n=1$ was proved in Ref.~\cite{crispino_2001}. 
We used the case $n=2$ in deriving Eq.~\eqref{eq:Tot_response_rate_final}.

We first establish some useful equalities.  By differentiating
the equation in Ref.~\cite[Eq.~8.824]{gradshteyn} we find
\begin{align}
    \dv[n+1]{\,}{z}Q_n(z)
    = \frac{(-1)^{n+1}2^n n!}{(z^2-1)^{n+1}}.\label{eq:diff-of-Q}
\end{align}
Then, for $n\geq 2$,
\begin{align}
    \dv[n+1]{\,}{z}Q_n(z) 
    - (2n-1)\dv[n]{\,}{z}Q_{n-1}(z)
    = \dv[2]{\,}{z}\left[\frac{(-1)^{n+1}2^{n-2}(n-2)!}{(z^2-1)^{n-1}}\right]. 
\end{align}
By integrating this equation twice and noting~\cite[Eq.~8.820.1]{gradshteyn}
\begin{align}
    Q_n(z) \approx \frac{n!}{(2n+1)!!}z^{-n-1},\ \ |z|\gg 1, \label{eq:large-z-Q}
\end{align}
we find
\begin{align}
     \dv[n-1]{\,}{z}Q_n(z) 
    = (2n-1)\dv[n-2]{\,}{z}Q_{n-1}(z)
    + \frac{(-1)^{n+1}2^{n-2}(n-2)!}{(z^2-1)^{n-1}}, \ \ n\geq 2.\label{eq:Q-Q-equation}
\end{align}

Next the $(n-1)$-th derivative of the Legendre equation satisfied by $Q_\ell(z)$ 
reads
\begin{align}
(z^2-1)\dv[n+1]{\,}{z}Q_\ell(z)
+ 2nz\dv[n]{\,}{z} Q_\ell(z)
& = [\Lambda_\ell - n(n-1)]\dv[n-1]{\,}{z}Q_\ell(z).
\label{to-be-proved}
\end{align}
From this equation we find
\begin{align}
\dv{\,}{z}\left\{
(z^2-1)^{2n}\left[\dv[n]{\,}{z}Q_\ell(z)\right]^2
\right\}
& = \left[\Lambda_\ell-n(n-1)\right](z^2-1)^{2n-1}
\dv{\,}{z}\left\{\left[ \dv[n-1]{\,}{z}Q_\ell(z)\right]^2\right\}. \label{eq-eq-eq}
\end{align}
This implies
\begin{align}
    \dv{\,}{z}\left[(z^2-1)^{2n}S_n(z)\right]
    = (z^2-1)^{2n-1}
    \dv{\,}{z}\left\{
    S_{n-1}(z)
    - \frac{2n-1}{[2(n-1)]!} \left[ \dv[n-1]{\,}{z}
    Q_{n-1}(z)\right]^2\right\}. 
\end{align}
The second term in curly brackets on the right-hand side subtracts
the $\ell=n-1$ term in $S_{n-1}(z)$.
Then, by using Eq.~\eqref{eq:diff-of-Q} with $n$ replaced by $n-1$ we obtain from
the above equation,
\begin{align}
    \dv{\,}{z}\left[(z^2-1)^{2n}S_n(z)\right]
    & = (z^2-1)^{2n-1}
    \dv{\,}{z}S_{n-1}(z)
    +\frac{(-1)^{n-1}2(2n-1)}{(2n-3)!!}(z^2-1)^{n-1}\dv[n-1]{\,}{z}
    Q_{n-1}(z)
    \,. \label{eq:intermediate}
\end{align}

Now, we prove Eq.~\eqref{eq:summation-formula} by induction. 
As we stated above, it is true for $n=1$. Assume
Eq.~\eqref{eq:summation-formula} with $n$ replaced by $n-1$, i.e.,
\begin{align}
    S_{n-1}(z) = (-1)^n \frac{2(n-1)}{(2n-3)!!}\frac{1}{(z^2-1)^n}
    \dv[n-2]{\,}{z}Q_{n-1}(z),\ \ n\geq 2.
\end{align}
By substituting this formula into Eq.~\eqref{eq:intermediate} we find
\begin{align}
     \dv{\,}{z}\left[(z^2-1)^{2n}S_n(z)\right]
    & = \dv{\,}{z}\left[ 
   (-1)^{n+1} \frac{2n}{(2n-3)!!}(z^2-1)^{n-1}
    \dv[n-2]{\,}{z}Q_{n-1}(z)\right].
\end{align}
Hence,
\begin{align}
    (z^2-1)^{2n}S_n(z)
    = (-1)^{n+1}\frac{2n}{(2n-3)!!}(z^2-1)^{n-1}\dv[n-2]{\,}{z}Q_{n-1}(z) + C_{n}\,, \label{eq:integrated-formula}
\end{align}
where $C_n$ is a constant, which we shall determine next.

The behavior of the $\ell$-th term in the series $S_n(z)$ for large $|z|$ 
can be found from Eq.~\eqref{eq:large-z-Q} as
\begin{equation}
    \left[ \dv[n]{\,}{z}Q_\ell(z)\right]^2 \sim z^{-2(\ell+n+1)},\ \ \ell\geq n.
\end{equation}
Hence, the left-hand side of Eq.~\eqref{eq:integrated-formula} tends to $0$ as
$z\to\infty$.  The $z\to\infty$ limit of the first term of this equation
can also be found using Eq.~\eqref{eq:large-z-Q}. Then, by requiring that the
right-hand side tends to $0$ as $z\to\infty$, we find
\begin{align}
    C_n & = \frac{2^{n-1} n(n-2)!}{(2n-1)!!}.
\end{align}
Hence,
\begin{align}
    S_n(z) & = (-1)^{n+1}\frac{2n}{(2n-1)!!}\frac{1}{(z^2-1)^{n+1}}
    \left[ (2n-1)\dv[n-2]{\,}{z}Q_{n-1}(z) 
    + \frac{(-1)^{n+1}2^{n-2}(n-2)!}{(z^2-1)^{n-1}}\right] \notag \\
    & = (-1)^{n+1}\frac{2n}{(2n-1)!!}\frac{1}{(z^2-1)^{n+1}}
    \dv[n-1]{\,}{z}Q_n(z)\,,
\end{align}
by Eq.~\eqref{eq:Q-Q-equation}.  Thus, Eq.~\eqref{eq:summation-formula} is true for all $n\geq 1$ by induction.

%%%%%%%%%%%%%%%%%%%%%%%%%%%%
%\section{The response rate with $t-$dependent stress-energy tensor}
%\label{sec:appendix_C}
%%%%%%%%%%%%%%%%%%%%%%%%%%%%
%...
%%%%%%%%%%%%%%%%%%%%%%%%%%%%
\section{Alternative stress-energy tensor and the response rate}
\label{sec:appendix_C}
%%%%%%%%%%%%%%%%%%%%%%%%%%%%
An alternative stress-energy tensor is 
constructed by choosing the following $rr$-component
\begin{equation}
    \label{eq:alpha_T_rr}
    \check{T}^{rr} = - \mu \frac{f(r)^{\alpha}}{r^2} \theta \left(r-r_0\right)\,\delta_\perp^{(2)},
\end{equation}
where $\mu, f(r) > 0$ and where $\alpha$ is a real parameter associated with the properties of the string. Substituting this expression into the conservation equation~\eqref{eq:r-comp_conserv_eq}, we obtain the corresponding $tt$-component,
\begin{equation}
    \label{eq:alpha_T_tt}
    \check{T}^{tt} = \mu \left[2\,\delta(r-r_0)\,f(r) + (2\alpha-1) \,\theta\left(r-r_0\right)\,f'(r) \right] \frac{f(r)^{\alpha-2}}{r^2 f'(r)}\, \delta_\perp^{(2)}.
\end{equation}
Here, $\mu$ is given by
\begin{equation}
    \label{eq:alpha_mu}
    \mu=\frac{1}{2}\,m_0\,f'(r_0)\,f(r_0)^{\tfrac{1}{2}-\alpha}.
\end{equation}
This stress-energy tensor satisfies the  
weak energy condition for $\alpha \geq 1.$ If the $tt$-component has only the point-mass contribution,  
i.e., if $\alpha=1/2$, then the corresponding stress-energy tensor violates the weak energy condition. The stress-energy tensor considered in Sec.~\ref{sec:response_rate_Unruh}, described by Eqs.~\eqref{eq:T_t_t}--\eqref{eq:T_r_r}, corresponds to the choice $\alpha=1$. 

Near the horizon and in Rindler coordinates, the stress-energy tensor $\check{T}^{\mu \nu}$ can be approximated as
\begin{align}
\check{T}^{\eta \eta} &\approx 
m_0\!\left[\delta(\xi) + (\!2\alpha-1) a\, \mathrm{e}^{-2(2-\alpha)a\xi} 
\theta(\xi)\right]\frac{1}{r_{h}^{2}}\,\delta_{\perp}^{(2)}, 
\\[4pt]
\check{T}^{\xi\xi} &\approx
- m_0 a \, \mathrm{e}^{-2(2-\alpha)a\xi}\,\theta(\xi)\,
\frac{1}{r_{h}^{2}}\,\delta_{\perp}^{(2)}.
\end{align}
By writing $\alpha = 1 - \beta/2$ and identifying $m_0 = \upmu$, we recover the same expression as that obtained in Rindler spacetime~\cite[Eqs.~(B8)--(B9)]{brito_2024_gw}. (Here $\beta$ and $\upmu$ denote the parameters used in Ref.~\cite{brito_2024_gw}.) In the main text of Ref.~\cite{brito_2024_gw} we have considered $\beta=0$, which corresponds here to
$\alpha=1$.

The response rate of the stress-energy tensor $\check{T}^{\mu \nu}$, given by Eqs.~\eqref{eq:alpha_T_rr}--\eqref{eq:alpha_T_tt}, to the Hawking radiation for fixed angular momentum number $\ell$, is derived by performing calculations similar to those in Sec.~\ref{sec:response_rate_Unruh}. The resulting response rate is given by
\begin{equation}
\label{eq:alpha_response_rate}
    \check{\mathcal{R}}_\ell = \frac{(2\ell+1)g_G^2m_0^2}{8\pi^2 k_\ell M} 
    \left[ r_0^2 f(r_0)^{5/4} \, \dv[2]{\,}{r_0} Q_\ell \left( \frac{r_0}{M}-1\right) + 2M \, (\alpha-1)\,r_0^{-2} f(r_0)^{1/4-\alpha}\int_{r_0}^{\infty} \dd r\, r^2 f(r)^\alpha \dv[2]{\,}{r} Q_\ell \left( \frac{r}{M}-1\right)\right]^2.
\end{equation}
If the second term on the right-hand side of this equation is interpreted as the contribution of the string to the response rate (in the 
specific gauge considered), this contribution vanishes for $\alpha = 1$. 

%%%%%%%%%%%%%%%%%%%%%%%%%%%%
\section{Total response rates for static classical point particle systems in static spacetimes}
\label{sec:appendix_D}
%%%%%%%%%%%%%%%%%%%%%%%%%%%%
Table~\ref{tab:formulas} summarizes the total response rates derived in this and previous works for classical 
static point particle systems acting as scalar sources $\sigma$, electric charges $q$ (and $\tilde{q}$ when interacting with a Proca field of mass $\tilde{\mathfrak{m}}$), or masses $m_0$  in static spacetimes. The listed results correspond to particles uniformly accelerating in Minkowski spacetime (equivalently, static in the Rindler wedge) with proper acceleration $a$, in the Minkowski vacuum, as well as to particles held static outside black holes or inside cosmological horizons at the radial coordinate position $r=r_0$ in a given quantum state, e.g., the Unruh or Hartle-Hawking vacuum states in Schwarzschild spacetime and Reissner-Nordström spacetime with charge $Q$ or Bunch-Davies vacuum state in de~Sitter spacetime with cosmological constant $\Lambda$. Recall that the proper acceleration $\varrho(r_0)$ is given by $\varrho \equiv \sqrt{\varrho^\mu \varrho_\mu}$, with $\varrho^\mu \equiv u^\nu \nabla_\nu u^\mu$.

\makeatletter
\setcounter{table}{0}
\renewcommand{\thetable}{\thesection\arabic{table}}
\makeatother
\begin{table}[t]
\centering
\renewcommand{\arraystretch}{2.5}
\begin{tabular}{c c c c}
\hline\hline
\textbf{Spacetime} 
& \textbf{Quantum state} 
& \textbf{Interaction} 
& \textbf{Total response rate} \\[0.3em]
\hline

\multirow{5}{*}{Minkowski / Rindler}
& \multirow{5}{*}{Minkowski vacuum}
& Scalar~\cite{ren_1994,higuchi_1997,higuchi_1998} 
& $\displaystyle \frac{\sigma^2 \, a}{4\pi^2}$ \\
& 
& Electromagnetic~\cite{higuchi_1992R,higuchi_1992,crispino_1998} 
& $\displaystyle \frac{q^2 \, a}{2\pi^2} \, \ln \left( \frac{a}{k_\perp^0}\right)$ \vspace{.1 cm}\\

& 
& Gravitational~\cite{brito_2024_gw} \textbf{(}see Eq.~\eqref{eq:Total_response_rindler_near_horizon}\textbf{)} 
& $\displaystyle \frac{m_0^2 \, a}{2\pi^2} \, \left( \frac{a}{k_\perp^0}\right)^2$ \vspace{.1 cm}\\

& 
&
Proca~\cite{castineiras_2011} & Given in Eq.~\eqref{eq:proca_tot_rate_2} \\

& 
&
Massive scalar~\cite{castineiras_2003_2} & Given in Eq.~\eqref{eq:massive_scalar_tot} \\
\hline

Minkowski ($\mathbb{M}^{p+2}$)
& 
Minkowski vacuum
&
Scalar~\cite{crispino_2004} & $\displaystyle \frac{\sigma^2 a^{p-1} \left[\left(\frac{p}{2}-1\right)!\right]^3}{4 \pi^{p/2 + 1} (p - 1)!}$ \vspace{.1 cm} \\
\hline

\multirow{6}{*}{Schwarzschild}
& \multirow{3}{*}{Unruh vacuum}
& Scalar~\cite{higuchi_1997,higuchi_1998}  
& $\displaystyle \frac{\sigma^2 \, \varrho(r_0)}{4\pi^2}$ \\

& 
& Electromagnetic~\cite{crispino_1998,crispino_2001} 
& $\displaystyle \frac{q^2 \, \varrho(r_0)}{2\pi^2} \, Q_1\left(\frac{2r_0}{r_h}-1 \right)$ \\

& 
& Gravitational %(Sec.~\ref{sec:response_rate_Unruh}) 
& $\displaystyle \frac{g_G^2 m_0^2\,\varrho(r_0)}{8\pi^2}\,\left[-\frac{2r_h}{3}\dv{\,\,}{r_0}Q_2 \left(\frac{2r_0}{r_h}-1 \right)\right]$ \vspace{.1 cm} \\

\cline{2-4}

& \multirow{3}{*}{Hartle-Hawking vacuum}
& Scalar~\cite{higuchi_1998} 
& $\displaystyle \frac{\sigma^2 \, \varrho(r_0)}{4\pi^2} + \frac{\sigma^2}{16\pi^2\,r_0^2\,\varrho(r_0)}$ \\

& 
& Electromagnetic (see Sec.~\ref{sec:hartle-hawking})
& $\displaystyle \frac{q^2 \, \varrho(r_0)}{2\pi^2} \, Q_1\left(\frac{r_0}{M}-1 \right)$ \\

& 
& Gravitational %(Sec.~\ref{sec:hartle-hawking}) 
& $\displaystyle \frac{g_G^2m_0^2\,\varrho(r_0)}{8\pi^2}\,\left[-\frac{2r_h}{3}\dv{\,\,}{r_0}Q_2 \left(\frac{2r_0}{r_h}-1 \right)\right]$ \vspace{.1 cm} \\

\hline

\multirow{3}{*}{Reissner-Nordström}
& Unruh vacuum
& \multirow{3}{*}{Scalar~\cite{castineiras_2000}} 
& $\displaystyle \frac{\sigma^2 \, \varrho(r_0)}{4\pi^2}\frac{ \left(M - \frac{Q^2}{r_h}\right)}{ \left(M - \frac{Q^2}{r_0}\right)}$ \vspace{.3 cm}\\
& Hartle-Hawking vacuum
& 
& \hspace{-1 cm}$\displaystyle \frac{\sigma^2 \, \varrho(r_0)}{4\pi^2}\frac{ \left(M - \frac{Q^2}{r_h}\right)}{ \left(M - \frac{Q^2}{r_0}\right)} + \frac{\sigma^2 \left(M - \frac{Q^2}{r_h}\right) \left(M - \frac{Q^2}{r_0}\right)}{4 \pi^2 r_h^2 r_0^2 \varrho(r_0)}$ \vspace{.1 cm}\\

\hline

de Sitter
& Bunch-Davies vacuum
& Scalar~\cite{castineiras_2003}  
& $\displaystyle \frac{\sigma^2}{4\pi^2 \sqrt{3/\Lambda}} \left( 1 + \frac{3}{\Lambda} \varrho(r_0)^2 \right)^{1/2}$  \vspace{.1 cm}\\

% \hline

% Schwarzschild-de Sitter
% & --
% & Scalar~\cite{castineiras_2003} 
% & No closed formula \\

\hline\hline
\end{tabular}
\caption{Response rate formulas for classical point particle systems interacting with bosonic fields in different spacetime backgrounds and quantum states.}\label{tab:formulas}
\end{table}

The electromagnetic and gravitational total response rates in Minkowski (or equivalently in Rindler) spacetime feature infrared divergences, as discussed previously. Accordingly, the corresponding total response rates are presented at leading order in an infrared cutoff imposed on the magnitude of the transverse momentum, $k_\perp^0$.

Note that, unlike the electromagnetic case, the total response rate for the Proca field in Rindler spacetime in the Minkowski vacuum is finite due to the presence of the mass of the vector field, which acts as a natural infrared cutoff and regulates the infrared behavior. To compute the the total response rate for the Proca field from the response rate with fixed transverse momenta derived in Ref.~\cite{castineiras_2011}, we proceed similarly to the electromagnetic and gravitational cases. The response rate for fixed transverse momentum for the Proca field is given by~\cite{castineiras_2011}
\begin{equation}
    \label{eq:proca_partial_rate}
    \mathcal{R}_{\mathbf{k}_\perp}^{\text{Proca}} = \frac{\tilde{q}^2}{4\pi^3\,a}\left|K_1\left(\frac{\sqrt{k_\perp^2 + \tilde{\mathfrak{m}}^2}}{a}\right)\right|^2.
\end{equation}
Then, we integrate over the transverse momenta, namely
\begin{eqnarray}
    \label{eq:proca_tot_rate_1}
\mathcal{R}_{\mathrm{tot}}^{\text{Proca}}
&=& 2\pi \int_{0}^{\infty} \mathrm{d}k_\perp\, k_\perp\,
\mathcal{R}_{\mathbf{k}_\perp}^{\text{Proca}} \\ \label{eq:proca_tot_rate_2}
&=& \frac{\tilde{q}^2}{4\pi^2}\frac{\tilde{\mathfrak{m}}^2}{a}\left[K_0\left(\frac{\tilde{\mathfrak{m}}}{a}\right)^2+2 \frac{a}{\tilde{\mathfrak{m}}} K_0\left(\frac{\tilde{\mathfrak{m}}}{a}\right)K_1\left(\frac{\tilde{\mathfrak{m}}}{a}\right)-K_1\left(\frac{\tilde{\mathfrak{m}}}{a}\right)^2\right],
\end{eqnarray}
which diverges in the limit $\tilde{\mathfrak{m}} \to 0.$ We expand $\mathcal{R}_{\mathrm{tot}}^{\text{Proca}}$ at leading order in $\tilde{\mathfrak{m}} \ll a$ and find
\begin{equation}
    \label{eq:proca_small_m}
\mathcal{R}_{\mathrm{tot}}^{\text{Proca}}
\approx \frac{\tilde{q}^2 \, a}{2\pi^2} \, \ln \left( \frac{a}{\tilde{\mathfrak{m}}}\right), \quad \text{for}\,\,\tilde{\mathfrak{m}} \ll a.
\end{equation}

The response rate for the scalar field with mass $\mathfrak{m}$ in Rindler spacetime is analyzed in Ref.~\cite{castineiras_2003_2}. The response rate for fixed transverse momentum is given by
\begin{equation}
    \label{eq:massive_scalar_parc}
    \mathcal{R}_{\mathbf{k}_\perp}^{\text{Scalar}(\mathfrak{m})} = \frac{\sigma^2}{4\pi^3\,a}\left|K_0\left(\frac{\sqrt{k_\perp^2 + \mathfrak{m}^2}}{a}\right)\right|^2.
\end{equation}
 The total response rate is obtained by integrating over the transverse momenta as
\begin{eqnarray}
    \label{eq:massive_scalar}
    \mathcal{R}_{\mathrm{tot}}^{\text{Scalar}(\mathfrak{m})}
&=& 2\pi \int_{0}^{\infty} \mathrm{d}k_\perp\, k_\perp\,
\mathcal{R}_{\mathbf{k}_\perp}^{\text{Scalar}(\mathfrak{m})} \\ \label{eq:massive_scalar_tot}
&=& \frac{\sigma^2}{4\pi^2}\frac{\mathfrak{m}^2}{a}\left[K_1\left(\frac{\mathfrak{m}}{a}\right)^2-K_0\left(\frac{\mathfrak{m}}{a}\right)^2\right],
\end{eqnarray}
which tends to the massless result, $\sigma^2 a/(4\pi^2)$, in the limit $\mathfrak{m} \to 0$.

\end{widetext}

%%%%%%%%%%%%%%%%%%%%%%%%%%%%%%%%%%%%

\end{document}